\documentclass[acmsmall,screen,nonacm]{acmart}

\acmJournal{PACMPL}
\setcopyright{none}

\usepackage{booktabs}
\usepackage{tabularx}
\usepackage{multirow}
\usepackage{xspace}
\usepackage{pifont}
\usepackage{capt-of}
\usepackage{enumitem}
\usepackage{tcolorbox}
\usepackage{tikz}
\usetikzlibrary{positioning,arrows.meta}

\newcommand{\toolname}{Sightline\xspace}
\newcommand{\Src}{S}
\newcommand{\pmax}{p_{\max}}

\newtcolorbox{answerbox}{colback=gray!5,colframe=gray!50,boxrule=0.5pt,left=3pt,right=3pt,top=2pt,bottom=2pt}
\newcommand{\cone}{\ding{182}\xspace}
\newcommand{\ctwo}{\ding{183}\xspace}
\newcommand{\cthree}{\ding{184}\xspace}
\newcommand{\leandecl}[1]{\texttt{#1}}
\theoremstyle{acmplain}
\newtheorem{assumption}{Assumption}

\begin{document}

\title{Sightline: Calibrated Data-Flow Summaries for LLM-Integrated Applications}

\author{Zihao Xu}
\affiliation{\institution{UNSW Sydney}\country{Australia}}
\affiliation{\institution{Southern Cross University}\country{Australia}}
\email{x123201298@gmail.com}

\author{Yuekang Li}
\affiliation{\institution{UNSW Sydney}\country{Australia}}
\email{yuekang.li@unsw.edu.au}

\author{Ruitao Feng}
\affiliation{\institution{Southern Cross University}\country{Australia}}
\email{ruitao.feng@scu.edu.au}

\author{Haifeng Shen}
\affiliation{\institution{Southern Cross University}\country{Australia}}
\email{haifeng.shen@scu.edu.au}

\begin{abstract}
Programs increasingly call large language models (LLMs), filling prompt placeholders with program values and consuming the output.
Such a call crosses the boundary between natural language and programming language (NL/PL).
To follow data across it, an analysis needs a data-flow summary of which placeholders the output uses, but no code describes what the model does.
Analyses of these LLM-integrated applications therefore taint every output with every input, creating spurious flows, or ask a second model to label each call, which is costly.
We prove, and mechanize in Lean~4, that a placeholder flows to the output of a transformer, to first order, only through the attention that reads it (no reading, no flow), and that a faint reading or sensitivity cannot certify that a placeholder does not flow.
Unlike reading, the log-probability gradients and Fisher signals of existing methods vanish on confident outputs, and single-span removal fails on duplicated content.
Our tool, \toolname, reads a recorded call in one forward pass, learns from labels which reads are uses, and calibrates the pruning to miss at most a chosen fraction $\alpha$ of true flows, on average over call sites, when deployment is exchangeable with calibration.
On 2,103 unseen repository families and four models, \toolname is the most accurate summary against the seven closest methods and a hidden-state probe.
Calibrated at $\alpha = 1\%$, it misses about 1\% of true flows and prunes 43.9--65.6\% of spurious flows, 1.6 to 2.6 times as many as the best baseline, and still 43.8--60.1\% when calibrated on another collection.
Through an open surrogate, it reads models treated as closed about as accurately as they read themselves, without their labels.
In the information-flow control of an LLM agent, it removes about half of the labels from inputs that do not flow.

\end{abstract}



\maketitle

\section{Introduction}
\label{sec:intro}

Large language models (LLMs) are now components of ordinary programs.
A program calls a model through an API at an \emph{LLM call site}.
It fills the placeholders of a prompt template with run-time values, such as a user query or a retrieved document, and then consumes the output, for example by storing it, sending it, or passing it to a tool.
Programs built this way are called LLM-integrated applications~\cite{liu2024llmsmith,xu2026prism}.

Program analyses that follow data, such as taint tracking, slicing and information-flow control~\cite{sabelfeld2003language}, must know at every call which inputs reach the output.
When the code of a callee is unavailable or too costly to analyze, they rely on a data-flow summary of it, as taint analyses do for framework code~\cite{arzt2014flowdroid,arzt2016stubdroid}.
LLM-integrated applications need these analyses as much as conventional software, because they pass untrusted input into prompts and act on the outputs, which has led to injection and code-execution vulnerabilities~\cite{liu2024llmsmith,he2026taintp2x}.

However, no code describes what happens inside an LLM call.
The model transforms the prompt in its weights rather than in program text, a crossing that prior work calls the boundary between natural language and programming language (NL/PL)~\cite{xu2026prism}.
For each call, an analysis needs to know which placeholders the output uses, and no summary answers this question.
In practice, on call sites from 2,103 open-source repository families (repositories with their forks), 12--14\% of the placeholders do not reach the output, because the model ignores them, lacks the context to use them, or refuses.
Every such placeholder that an analysis assumes to flow spreads a label to data that the output never used (label creep~\cite{sabelfeld2003language}).
A useful summary must miss few true flows, must prune the placeholders that do not flow, must be cheap enough to compute for every call, and must state how often it may miss.
This missing summary at the NL/PL boundary is the gap we address.

\begin{figure}[t]
\centering
\includegraphics[width=\linewidth]{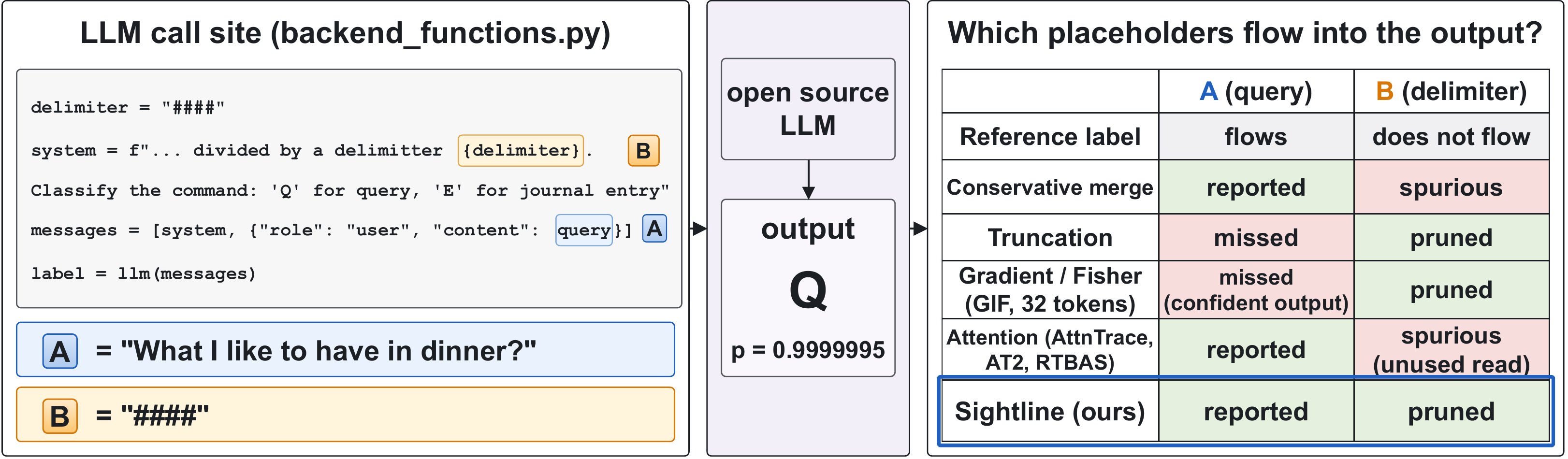}
\Description{A call site with two placeholders, the query A and the delimiter B, the confident output Q, and a grid with the verdict of each method on both placeholders.}
\caption{A call site from the new batch. The query (A) decides the confident output \texttt{Q}, and the delimiter (B) is read but unused. Each cell shows the verdict of a method at its threshold, green if right and red if wrong.}
\label{fig:example}
\end{figure}

\autoref{fig:example} shows the problem on one call site with two placeholders.
The output is a short decision that the first placeholder determines and that the model emits with near certainty.
The model also reads the second placeholder, but its output does not use it.
A correct summary thus reports the first placeholder and prunes the second.

Existing techniques do not produce this summary.
Program analyses either treat the output of an LLM call as opaque, which truncates every flow through it, or taint it as a whole, as a taint source~\cite{he2026taintp2x} or by merging the labels of all inputs into it~\cite{costa2025fides}.
Conservative merge never misses a flow, but labels then spread to data that the output never used.
Black-box per-call labeling asks LLM labelers to read the prompt and the output~\cite{xu2026prism}.
It judges each call directly, but one labeler costs about 10k tokens and 13 seconds per placeholder (\S\ref{sec:rq4}) and runs offline.
White-box methods use the computation of the model instead.
Context attribution removes parts of the input, as ContextCite and TracLLM do~\cite{cohenwang2024contextcite,wang2025tracllm}, sometimes fitting a linear model to the effects, or reads attention, as AttnTrace and AT2 do~\cite{wang2026attntrace,cohenwang2025at2}.
White-box information-flow control bounds the flow with gradients and Fisher information, as Geometric Information Flow (GIF) does~\cite{storek2026gif}, or screens dependencies with attention, as RTBAS does~\cite{zhong2025rtbas}.
Gradient and Fisher sensitivity carry a factor $1-p$, where $p$ is the probability of the output token, so they vanish when the model is confident, and removal fails when content is duplicated (\S\ref{sec:p4}).
Confident decisions are common at call sites, since for one in five outputs of Qwen3-8B~\cite{yang2025qwen3} the model leaves less than 1\% of probability to other tokens.
On the call of \autoref{fig:example}, the gradient- and Fisher-based methods miss the first placeholder for this reason.
The attention-based methods avoid this factor but report the second placeholder, because a model can read an input without using it.
None of them comes with a miss guarantee per call site.

Building such a summary is hard for three reasons.
\cone~\emph{The callee has no code.} Program analyses derive summaries from the code of the callee, so at an LLM call they fall back to truncation or conservative merge. It is unclear which evidence about the computation of a model can bound the flow at all.
\ctwo~\emph{Models read more than they use.} Like the delimiter of \autoref{fig:example}, a placeholder can be read but not used. Reporting every placeholder that the model reads thus also reports many placeholders that do not flow. Measuring how much the output changes when a placeholder is removed or perturbed reports fewer, but it misses flows when the model is confident or when its content is duplicated.
\cthree~\emph{A faint signal can hide a true flow.} In code, an analysis prunes an input safely when no dependence connects it to the result. At an LLM call, attention reads every placeholder at least a little, and a placeholder that is read only faintly, or whose removal changes nothing, can still decide the output. Any summary that prunes may therefore miss a flow, and an analysis can rely on it only if the summary also states how often it misses.

To address these challenges, we present \toolname, which summarizes a recorded call by reading the attention of the model.
Although the model has no program text, its computation has a fixed structure, and that structure decides where information can travel.
\textbf{In a transformer, a placeholder flows to the output, to first order, only through the attention that reads it, hence no reading, no flow.}
For \cone, we prove this property for every layer, in the way taint propagates along dependence edges in code, with attention as the edges.
This tells which evidence to read, and \toolname records in one forward pass how much the output reads each placeholder, while the guarantee of the summary comes from calibration.
For \ctwo, we prove that a head can read a placeholder heavily without using it, so \toolname learns from labeled calls which reads are uses, with a pruning readout over the reading of every attention head.
For \cthree, we prove that faint reading, faint gradients and removal without effect can each occur on a placeholder that decides the output, so \toolname states how often it misses.
It sets its threshold from labeled call sites of other repositories with conformal prediction~\cite{vovk2022alrw,dunn2023hierarchical}, so that, when new call sites come from the same distribution as these, a true flow of a randomly drawn call site is missed with probability at most a chosen $\alpha$, such as 1\%.

We fit \toolname on an old batch of 8,105 placeholder--output pairs and evaluate it on 19,672 new pairs from 2,103 repository families that training never saw, on four open models including a Mamba2 hybrid~\cite{dao2024ssd}.
\toolname is the most accurate summary on all four models (81.1--86.7 Macro-F1, the mean F1 score of the two classes), significantly so on the three transformers, where it leads the best of the seven closest methods by 4.5 to 6.2 points.
Given the same labels and call features (lengths and placeholder count), all seven baselines and a hidden-state probe stay significantly behind, so the precision comes from the readout on reading.
It holds up where the existing signals fail, with an area under the ROC curve (AUROC) of 99.0--100.0 on duplicated content, against 54.0--68.8 for leave-one-out removal (LOO), and a lead of 19 to 34 points over the Fisher signal on the most confident outputs.
Calibrated to miss about one true flow in a hundred, it prunes 43.9--65.6\% of the spurious flows that conservative merge reports, 1.6 to 2.6 times as many as the best baseline, and still 43.8--60.1\% when calibrated on the old batch.
It costs one forward pass per call (0.12--0.20\,s), and since reading needs only the recorded tokens, an open surrogate reads models treated as closed about as accurately as they read themselves, without their labels.
In the information-flow control of an LLM agent, it removes about half of the label creep and lets no attack call through an integrity policy, which blocks side-effecting calls that carry untrusted labels.
Together, these results show that reading carries program analyses across the NL/PL boundary with a summary that is accurate, calibrated per call site, and cheap enough for every call.

In summary, we make the following contributions.
\begin{itemize}[leftmargin=*]
\item \textbf{What one forward pass can certify about flow through an LLM call.}
Relative to the information-flow semantics of GIF, we show that attention routes soundly over-approximate local flow (\S\ref{sec:summaries}) through an LLM call.
We also show that a faint reading or sensitivity cannot certify that a placeholder does not flow.
All thirteen results are mechanized in Lean~4~\cite{demoura2021lean4} (\S\ref{sec:mechanization}), and \S\ref{sec:evaluation} confirms their predictions.
\item \textbf{Calibrated data-flow summaries across the NL/PL boundary.}
To the best of our knowledge, \toolname is the first white-box data-flow summary for LLM calls whose miss rate is calibrated per call site, and through an open surrogate it reads models treated as closed about as accurately as they read themselves, without their labels (\S\ref{sec:approach}).
\item \textbf{Evidence on unseen programs.}
To the best of our knowledge, this is the first evaluation of white-box flow signals on real LLM call sites of open-source programs.
On repositories that training never saw and on four models, \toolname is more accurate than the seven closest methods, holds up where the theory predicts that the existing signals fail, and helps a downstream information-flow client (\S\ref{sec:evaluation}).
\end{itemize}

\section{Background}
\label{sec:problem}

This section defines the call sites we summarize, what a correct summary is, and the transformer notation that the theory in \S\ref{sec:theory} uses.

\subsection{LLM Call Sites}
\label{sec:callsites}

An \emph{LLM call site} is a program point that sends messages to a language model.
The messages come from a prompt template, and the template contains \emph{placeholders} that the program fills with run-time values, such as a user query, a retrieved document, or a conversation history.
Each call renders the template into a prompt, which the model reads as tokens $x_1,\dots,x_P$.
Each placeholder then occupies a span of positions $\Src \subseteq \{1,\dots,P\}$, and the model produces the output tokens $y_1,\dots,y_n$.
We call the pair of the rendered prompt and the output that the model actually produced a \emph{recorded call}.

A data-flow summary of the call tells the consuming analysis which placeholders the output carries.
Runtime taint tracking and information-flow control propagate labels through the call with it~\cite{costa2025fides,debenedetti2025camel}, and offline analyses use it to follow data through recorded calls.
Because whether the model uses a placeholder depends on the values and on what the model generates, we compute the summary for each recorded call, as dynamic taint tracking does for each execution of ordinary code.

\subsection{Problem Formalization}
\label{sec:summaries}

To judge a summary, we need a notion of flow that matches what the consuming analysis cares about, namely whether the output carries the content of a placeholder or a decision derived from it.
We adopt the notion of reach of prior black-box summaries of LLM calls~\cite{xu2026prism}.

\begin{definition}[Reach]
\label{def:reach}
A placeholder $\Src$ of a recorded call \emph{reaches} the output $y$ if $y$ visibly reflects the content of $\Src$, namely if $y$ copies or transforms that content (for example, by rewriting, translating, or summarizing it) or if the content determines a decision, a selection, a computed value, the format, or a quantitative property of $y$.
Otherwise, $\Src$ does not reach $y$.
This is the case when $y$ shows no observable use of $\Src$, for instance because the model ignored it, lacked the context or the capability to use it, or refused the request.
\end{definition}

\noindent
Reach is a semantic property of one recorded call, and it is what black-box per-call labeling decides when LLM labelers read the prompt, the placeholder and the output~\cite{xu2026prism}.
We write $r(\Src) \in \{0,1\}$ for the reference label, and we call a placeholder with $r(\Src)=1$ a \emph{true flow} and one with $r(\Src)=0$ a \emph{non-flow}.
The theory in \S\ref{sec:theory} also uses \emph{local flow}, the sensitivity of the output to small perturbations of the placeholder and the mutual information that such perturbations carry to the next token~\cite{storek2026gif}.
The two notions differ, since a confident decision can reach the output while its local flow is almost zero (\S\ref{sec:p4}), and we state for each result which notion it concerns.
For an analysis that propagates labels, reach is the notion that matters, because a decision derived from a placeholder must carry its label even when the model was certain.

\begin{definition}[Summary]
\label{def:summary}
A \emph{summary} of a recorded call reports a subset of its placeholders as flowing to the output and prunes the others.
A \emph{miss} is a true flow that the summary prunes.
A \emph{spurious flow} is a non-flow that the summary reports.
The \emph{pruning ratio} of a summary is the fraction of non-flows that it prunes.
\end{definition}

\noindent
The two summaries that analyses use today sit at the extremes of this definition.
Conservative merge reports every placeholder, so it never misses a flow but has a pruning ratio of zero.
Truncation reports nothing, so it prunes every non-flow and misses every true flow.
A useful summary must prune non-flows while missing few true flows (challenges \cone and \ctwo).

A summary that decides from evidence will sometimes err (challenge \cthree), so we ask for a bound on its misses.
The bound should hold at the granularity at which an analysis consumes summaries, namely the call site.

\begin{definition}[$\alpha$-calibration per call site]
\label{def:alpha}
A summary is \emph{$\alpha$-calibrated per call site} if, for a call site drawn from the deployment distribution and a true flow drawn uniformly from the true flows of its recorded calls, the probability that the summary misses the flow is at most $\alpha$, where the probability also ranges over the data on which the summary was calibrated.
\end{definition}

\noindent
We state the bound per call site rather than per pair because the true flows of one call site share a template and tend to be missed together, so a rate averaged over pairs can be dominated by a few call sites.

\paragraph{Scope}
We summarize one recorded call at a time, which is what runtime monitors and offline analyses consume, and these analyses compose the summaries along the program and across the steps of an agent.

\subsection{Attention in Transformers}
\label{sec:transformer}

The theory in \S\ref{sec:theory} relies on one structural fact of the transformer~\cite{vaswani2017attention}, so we fix the notation here for the decoder-only models we study.
To read a recorded call, we run the model once on $[x_1,\dots,x_P,y_1,\dots,y_{n-1}]$ with teacher forcing, so every position sees exactly the tokens of the recorded call.
The \emph{output-predicting positions} $Q = \{P,\dots,P+n-1\}$ are the positions whose next token is an output token.
A model with $L$ layers of $H$ heads keeps a residual stream $h^{\ell}_j \in \mathbb{R}^{d_{\mathrm{model}}}$ at layer $\ell$ and position $j$.
In a pre-norm layer, head $h$ normalizes the stream, $\hat{h}_i = \mathrm{Norm}(h^{\ell}_i)$, and computes queries $q_t = W_Q \hat{h}_t$, keys $k_i = W_K \hat{h}_i$ and values $v_i = W_V \hat{h}_i$ of dimension $d$.
Its attention weights and output at position $t$ are
\begin{equation}
\label{eq:attention}
a^{\ell h}_{ti} = \operatorname{softmax}_{i \le t}\!\left(\frac{q_t^{\top} k_i}{\sqrt{d}}\right),
\qquad
\bar{o}_t = \sum_{i \le t} a^{\ell h}_{ti}\, v_i,
\qquad
o_t = W_O\, \bar{o}_t ,
\end{equation}
and the layer adds the outputs of all heads to the stream before a position-wise multilayer perceptron (MLP) block.
Normalization and MLPs act on each position separately.
Attention is therefore the only operation that moves information from one token position to another, the view on which circuit analyses of transformers build~\cite{elhage2021framework}.
Sliding-window layers~\cite{beltagy2020longformerlongdocumenttransformer} set the weights outside their window to zero, and hybrid models replace some attention layers with state-space layers, which also move information across positions.
We treat hybrid models in \S\ref{sec:hybrid}.

\section{Four Properties of Reading}
\label{sec:theory}

We call the attention that later positions of a recorded call pay to a placeholder span the \emph{reading} of that span.
The output-predicting positions read the span directly, and for head $(\ell,h)$ and position $t \in Q$, the reading mass is
\begin{equation}
\label{eq:mass}
m^{\ell h}_t(\Src) \;=\; \sum_{i \in \Src} a^{\ell h}_{ti}.
\end{equation}
This section asks what one forward pass of a recorded call can certify about flow, and answers with four properties of reading.
All results of this section are mechanized in Lean~4~\cite{demoura2021lean4} with Mathlib~\cite{mathlib2020} (\S\ref{sec:mechanization}), and the proofs below sketch the mechanized ones.
Reading is sound, since relative to the information-flow semantics of GIF~\cite{storek2026gif} it bounds local flow by reading masses scaled by local gains (\textbf{P1}, \S\ref{sec:p1}).
Reading over-reports, and a faint reading cannot certify a non-flow, since reading can vanish on a flow (\textbf{P2}, \S\ref{sec:p2}), so the guarantee of a pruning decision is statistical, and calibration over clusters of call sites provides it for any score (\textbf{P3}, \S\ref{sec:p3}).
Sensitivity evidence certifies no more, since gradient and Fisher signals collapse on confident outputs and single-span removal collapses on duplicated content, and this collapse predicts which scores can prune under the guarantee (\textbf{P4}, \S\ref{sec:p4}).
The properties also motivate three choices of \toolname, namely to read the span directly in every head (\autoref{thm:route}), to prune with a readout that is not monotone in the reading masses (Proposition~\ref{prop:monotone}), and to calibrate a score without a $(1-p)$ factor (Corollary~\ref{cor:threshold}).
Each property makes a prediction that \S\ref{sec:evaluation} tests.
Reading reports every flow in which the output copies a value that we planted, and a monotone readout loses precision (\S\ref{sec:rq1}), sensitivity collapses on confident outputs and removal on duplicated content (\S\ref{sec:rq2}), and every signal misses about $\alpha$ while sensitivity signals prune little (\S\ref{sec:rq3}).
\S\ref{sec:hybrid} treats hybrid models.

\subsection{P1: Soundness of Reading}
\label{sec:p1}

The callee has no code (challenge \cone), but its computation has a fixed structure, so we first ask how a change at one position can affect a later position through a single head.
Write $\hat{h}_i = \mathrm{Norm}(h^{\ell}_i)$ for the normalized stream that a head of layer $\ell$ reads at position $i$.

\begin{lemma}[Routing bound of one head]
\label{lem:head}
For a head with the notation of \autoref{eq:attention} and positions $s < t$,
\[
\frac{\partial o_t}{\partial \hat{h}_s} \;=\; a_{ts}\, W_O \left[ W_V + \frac{(v_s - \bar{o}_t)\, q_t^{\top} W_K}{\sqrt{d}} \right],
\qquad\text{hence}\qquad
\left\| \frac{\partial o_t}{\partial \hat{h}_s} \right\|_2 \le a_{ts}\, c_{ts},
\]
where $c_{ts} = \|W_O\|_2 \big( \|W_V\|_2 + \|v_s - \bar{o}_t\|_2 \|q_t\|_2 \|W_K\|_2 / \sqrt{d} \big)$.
\end{lemma}

\begin{proof}[Proof sketch]
Since $q_t$ depends on position $t$ alone, $\hat{h}_s$ enters $o_t$ only through $v_s$, which contributes $a_{ts} W_O W_V$, and through $k_s$, where $\partial a_{ti} / \partial k_s = a_{ti}(\delta_{is} - a_{ts})\, q_t^{\top}/\sqrt{d}$ yields $a_{ts} W_O (v_s - \bar{o}_t) q_t^{\top} W_K / \sqrt{d}$.
Submultiplicativity gives the bound.
Query and key normalization, as used in Qwen3~\cite{yang2025qwen3}, and rotary embeddings~\cite{Su_2024} only multiply $W_K$ by a local Jacobian or a rotation, absorbed into $c_{ts}$, and grouped-query attention~\cite{ainslie-etal-2023-gqa} leaves the bound per head unchanged.
\end{proof}

\noindent
Both channels through which position $s$ can influence the output of a head, its value and its key, carry the factor $a_{ts}$.
A head that does not attend to $s$ cannot pass a first-order change at $s$ to $t$.

We next compose heads across layers, under the structure that \S\ref{sec:problem} fixes.

\begin{assumption}[Causal layers]
\label{asm:layers}
Along the forward pass of the recorded call, every layer is differentiable, normalization and MLPs act on each position separately, and a position influences another position of the same layer only through the heads of that layer.
\end{assumption}

\noindent
Decoder-only transformers satisfy Assumption~\ref{asm:layers} by construction, and Lean derives it for every stack of causal pre-norm layers whose components are differentiable (\S\ref{sec:mechanization}).
We treat the hybrid models that do not satisfy it in \S\ref{sec:hybrid}.
Let $x_\Src$ collect the input embeddings of the span $\Src$.
Define $\theta^0_j = 1$ if $j \in \Src$ and $0$ otherwise, and
\begin{equation}
\label{eq:taint}
\theta^{\ell+1}_j \;=\; \Lambda^{\ell}_j\, \theta^{\ell}_j \;+\; \rho^{\ell}_j \sum_{i<j} \sum_{h} a^{\ell h}_{ji}\, c^{\ell h}_{ji}\, \theta^{\ell}_i ,
\end{equation}
where $\Lambda^{\ell}_j$ bounds the Jacobian of layer $\ell$ at position $j$ with respect to its own stream (residual connection, the head channels of position $j$, normalization and MLP), $\rho^{\ell}_j$ bounds the Jacobian of the residual-plus-MLP map that follows attention, and $c^{\ell h}_{ji}$ is the constant of Lemma~\ref{lem:head} multiplied by the norm of the normalization Jacobian at $i$.

\begin{theorem}[Routing bound]
\label{thm:route}
Under Assumption~\ref{asm:layers}, $\big\| \partial h^{\ell}_t / \partial x_\Src \big\|_2 \le \theta^{\ell}_t$ for every layer $\ell$ and position $t$, and the Jacobian is zero when no attention route leads from $\Src$ to $t$.
\end{theorem}

\begin{proof}[Proof sketch]
By induction on $\ell$, from the selection $\partial h^{0}_t / \partial x_\Src$ of norm $1[t \in \Src]$.
The chain rule splits $\partial h^{\ell+1}_t / \partial x_\Src$ over the positions $i \le t$.
The term $i = t$ is at most $\Lambda^{\ell}_t \theta^{\ell}_t$, and by Assumption~\ref{asm:layers} every $i < t$ acts only through the heads, which Lemma~\ref{lem:head} bounds by $a^{\ell h}_{ti} c^{\ell h}_{ti}$, followed by the residual-plus-MLP map ($\rho^{\ell}_t$).
The same induction gives $\theta^{\ell}_t = 0$ when no route leads from $\Src$ to $t$.
\end{proof}

\noindent
\autoref{eq:taint} is quantitative taint propagation over the attention graph, in which a position becomes tainted only through edges with attention weight and the taint it receives is scaled by that weight.
Every route from $\Src$ starts where a later position reads $\Src$, so a placeholder that no later position reads has no route and, locally, no flow.
The bound covers every route from $\Src$ to $t$, including routes that pass through intermediate positions, which language models can use~\cite{ferrando2024routes}.
Its term with a single cross-position hop is the direct route from $\Src$ to an output-predicting position within one layer, which is the reading mass $m^{\ell h}_t(\Src)$ of \autoref{eq:mass} scaled by local gains.
\toolname reads this direct route (\S\ref{sec:approach}).
On Qwen3-8B the direct route carries most of the flow, a median of 71\% of the effect of removing the placeholder, and adding the two-hop routes to the features does not improve Macro-F1.
P2 and P3 turn the reading that P1 motivates into a pruning rule whose guarantee is statistical.

Theorem~\ref{thm:route} bounds sensitivity, and an information-theoretic notion of flow inherits the bound.
GIF~\cite{storek2026gif} defines the flow from a span to the next token $Y_t$ as the mutual information $I(U_\Src; Y_t)$ when the embeddings of the span receive an isotropic Gaussian perturbation $U_\Src$ of variance $\sigma^2$.
It proves that this flow is at most $\tfrac{1}{2}\log\det(I + \sigma^2 M_t) + o(\sigma^2)$, where $M_t = J_t^{\top} W^{\top} F_t W J_t$, $J_t$ is the Jacobian of the final hidden state with respect to the span, $W$ the unembedding, and $F_t = \operatorname{diag}(p) - p p^{\top}$ the Fisher matrix of the output distribution $p$.

\begin{assumption}[Local regime of GIF]
\label{asm:gif}
The map from the span embeddings to the final hidden state is continuous, differentiable at the embeddings of the recorded call with a local quadratic remainder bound, and $U_\Src$ is a centered isotropic Gaussian perturbation, as in the bound of GIF.
\end{assumption}

\begin{corollary}[Local soundness of routing]
\label{cor:gif}
Under Assumptions~\ref{asm:layers} and~\ref{asm:gif}, $I(U_\Src; Y_t) \le \tfrac{\sigma^2}{2}\, \kappa_t\, (\theta^L_t)^2 + o(\sigma^2)$, with $\kappa_t = \|J^{\mathrm{norm}}_t\|_2^2\, \|W\|_2^2\, \|F_t\|_2\, |\Src|\, d_{\mathrm{model}}$ and $J^{\mathrm{norm}}_t$ the Jacobian of the final normalization.
In particular, if no attention route leads from $\Src$ to $t$, the local flow from $\Src$ to $Y_t$ vanishes.
\end{corollary}

\begin{proof}[Proof sketch]
Use $\log\det(I+\sigma^2 A) \le \sigma^2 \operatorname{tr} A$ for positive semidefinite $A$, then $\operatorname{tr}(J^{\top} A J) \le \|A\|_2 \|J\|_F^2$ and $\|J\|_F^2 \le |\Src|\, d_{\mathrm{model}}\, \|J\|_2^2$, and finally $\|J_t\|_2 \le \|J^{\mathrm{norm}}_t\|_2\, \theta^L_t$ by \autoref{thm:route}.
\end{proof}

\noindent
Corollary~\ref{cor:gif} is local, like the guarantee of GIF.
Its constant contains $\|F_t\|_2$, which vanishes as the model becomes confident (\S\ref{sec:p4}), so local flow and reach part ways on confident outputs.
The guarantee for reach therefore comes from calibration on reach labels (\S\ref{sec:p3}).
P1 is a claim about flow rather than importance, which the debate on attention as explanation concerns~\cite{jain2019attention,wiegreffe2019attention}.

\subsection{P2: Incompleteness of Reading}
\label{sec:p2}

Models read more than they use (challenge \ctwo), so a sound over-approximation is only useful if its imprecision can be removed.
We next ask whether reading can also bound flow from below.

\begin{proposition}[Reading gives no lower bound]
\label{prop:incomplete}
For every attention pattern, including one that puts almost all weight on $s$, there are heads with that pattern for which $\partial o_t / \partial \hat{h}_s = 0$.
Consequently, a lower bound on local flow that depends only on reading masses must be zero.
\end{proposition}

\begin{proof}[Proof sketch]
If the output-value map $W_O W_V$ is zero, both terms of Lemma~\ref{lem:head} vanish whatever the attention, since $W_O(v_s - \bar{o}_t) = W_O W_V(\hat{h}_s - \sum_i a_{ti}\hat{h}_i)$, and a head with the same attention and the negated map cancels any head exactly.
Equal reading masses thus come with and without flow.
\end{proof}

\noindent
Conversely, a small reading at the recorded input does not show that a placeholder does not flow, because the content of the placeholder also sets its key and can raise its own reading.

\begin{proposition}[A vanishing read is not the absence of flow]
\label{prop:vanish}
There are heads in which the reading mass of a placeholder at the recorded input is arbitrarily small while a finite change of the placeholder flips the output.
\end{proposition}

\begin{proof}[Proof sketch]
Let a head attend to the placeholder at position $s$ and to a sink position, with key logits $\beta u$ and $0$ for the value $u$ at $s$, and let the values at $s$ and at the sink favor two different output tokens.
At the recorded $u = -1$, the reading $a_{ts} = \sigma(-\beta)$ tends to $0$ as $\beta \to \infty$, while at $u = 1$ it tends to $1$, and the greedy output switches.
For $\beta = 20$, the recorded reading is $2.1 \times 10^{-9}$.
\end{proof}

\noindent
This is the attention analogue of an implicit flow through a branch not taken~\cite{denning1976lattice}.
In our data no true flow is read this little, since each of the 11,746 true flows of Qwen3-8B is read at $10^{-2}$ or more by at least 11 heads, and by over 1,000 of the 1,152 heads at the median.
By Propositions~\ref{prop:incomplete} and~\ref{prop:vanish}, neither a large nor a small reading certifies whether a placeholder flows.
Precision must come from a second component that decides which reads are uses.
Information-flow analysis has the same structure, in which a sound tracker over-approximates and raises false alarms~\cite{sabelfeld2003language,hammer2009flow} and a separate step prunes them, as taint analyses of LLM-integrated applications do with an LLM~\cite{he2026taintp2x}.
We instead learn a \emph{pruning readout}, or readout for short, from the reading itself (\S\ref{sec:readout}).

\subsection{P3: Statistical Guarantee}
\label{sec:p3}

To address challenge \cthree, we ask what form the guarantee of a readout can take, given that it should prune unused reads without pruning true flows.
Write $\mu(\Src) = (\mu_e(\Src))_e$ for the reading masses of a span, one per head $e$, averaged or maximized over the output-predicting positions.
The direct-route term of \autoref{thm:route} is bounded by $\sum_e g_e\, \mu_e(\Src)$ with nonnegative gains $g_e$, so scores that are monotone in every $\mu_e$ have the same shape as the routing bound.

\begin{proposition}[Monotone scores cannot separate some unused reads]
\label{prop:monotone}
(a) A score that is non-decreasing in every $\mu_e$ never ranks a span above another span whose reading masses are at least as large in every head.
(b) There are spans $\Src_0, \Src_1$ of one call with $\mu(\Src_0) \ge \mu(\Src_1)$ in every head such that $\Src_0$ does not flow and $\Src_1$ flows.
Every monotone score then ranks $\Src_0$ at or above $\Src_1$, so no threshold reports $\Src_1$ while pruning $\Src_0$.
\end{proposition}

\begin{proof}[Proof sketch]
Part (a) follows by increasing one coordinate at a time.
For (b), let positional terms fix the attention, as with relative position biases and saturated logits, so that head $A$ puts weight $0.6$ on $\Src_0$ and $0.4$ on $\Src_1$, and head $C$ attends to $\Src_0$ and writes $-0.6$ times the output-value map of $A$.
Every head reads $\Src_0$ at least as much as $\Src_1$, but only $\Src_1$ changes the output.
\end{proof}

\noindent
A score that prunes such unused reads must be non-monotone in some heads, and it can then invert the order of the routing bound, so the bound no longer certifies its decisions.
We obtain the guarantee from statistics instead.

Split conformal prediction gives finite-sample guarantees for any score~\cite{vovk2022alrw,angelopoulos2023gentle}.
A per-pair guarantee assumes that pairs are exchangeable, which fails when the flows of one call site are missed together.
We therefore calibrate at the level of clusters of call sites, following conformal methods for two-level hierarchical data~\cite{dunn2023hierarchical,lee2023hierarchical}.

\begin{assumption}[Exchangeable clusters]
\label{asm:clusters}
The calibration clusters $G_1,\dots,G_{n_c}$ and a new cluster $G_{n_c+1}$, each containing at least one true flow, are exchangeable, the flows within a cluster may be arbitrarily dependent, and one true flow is drawn from each cluster by the same rule, independently across clusters.
\end{assumption}

\begin{theorem}[Per-call-site calibration]
\label{thm:conformal}
Under Assumption~\ref{asm:clusters}, let $s_i$ be the score of the flow drawn from $G_i$ and $k = \lfloor \alpha (n_c+1) \rfloor$, let $\tau_\alpha$ be the $k$-th smallest of $s_1,\dots,s_{n_c}$ (or $-\infty$ if $k=0$), and report a placeholder as flowing if and only if its score is at least $\tau_\alpha$.
Then $\Pr[s_{n_c+1} < \tau_\alpha] \le k/(n_c+1) \le \alpha$.
\end{theorem}

\begin{proof}[Proof sketch]
Extending the subsampling argument of \citet{dunn2023hierarchical} from independent and identically distributed groups to exchangeable clusters, independent draws make $s_1,\dots,s_{n_c+1}$ exchangeable.
At most $k$ of the $n_c+1$ scores lie strictly below the $k$-th smallest of the others, so by exchangeability $s_{n_c+1}$ does with probability at most $k/(n_c+1)$, with no tie-breaking.
\end{proof}

\noindent
In our experiments, the clusters are repository families, the unit by which our data are split, which treats all call sites of a family as dependent.
Each draw picks, at random, one of the call sites of the family that have a true flow and then one of its true flows, so \autoref{thm:conformal} gives the calibration of Definition~\ref{def:alpha} for call sites drawn through their families.
\S\ref{sec:rq3} measures this miss rate per call site on families that calibration never saw, from the same collection and from another one.

\subsection{P4: Collapse of Sensitivity}
\label{sec:p4}

Existing white-box methods judge flow by sensitivity, removing or perturbing the placeholder and measuring how much the output changes, or differentiating the output with respect to the placeholder.
We ask when this evidence fails.
Consider an output-predicting position with logits $z = W\, \mathrm{Norm}(h^L_t)$, output distribution $p = \operatorname{softmax}(z)$ and recorded token $y$.
Let $u$ be the embeddings of the placeholder and $J = \partial z / \partial u$.

\begin{proposition}[Saturation bounds sensitivity]
\label{prop:saturation}
(a) $\|\nabla_u \log p_y\|_2 \le \sqrt{2}\,(1 - p_y)\, \|J\|_2$, so gradient-based attributions vanish as $p_y \to 1$ while $\|J\|$ stays bounded.
(b) $\operatorname{tr}(J^{\top} F J) \le 2 (1 - \pmax)\, \|J\|_F^2$ with $F = \operatorname{diag}(p) - p p^{\top}$, so the Fisher influence and capacity of GIF vanish as $\pmax \to 1$ while $\|J\|$ stays bounded.
(c) If $p_y \ge m$ both with and without the placeholder, then removing the placeholder changes $\log p_y$ by at most $(1-m)/m$.
\end{proposition}

\begin{proof}[Proof sketch]
(a) $\nabla_z \log p_y = e_y - p$ and $\|e_y - p\|_2^2 = (1-p_y)^2 + \sum_{c \ne y} p_c^2 \le 2(1-p_y)^2$.
(b) $F$ is positive semidefinite with $\lambda_{\max}(F) \le \operatorname{tr} F = 1 - \sum_c p_c^2 \le 1 - \pmax^2 \le 2(1-\pmax)$, and $\operatorname{tr}(J^{\top} F J) \le \lambda_{\max}(F) \|J\|_F^2$.
(c) Both log-probabilities lie in $[\log m, 0]$, and $-\log m \le (1-m)/m$.
\end{proof}

\noindent
A vanishing local signal would be harmless if it only occurred when the placeholder does not matter.
The next result shows that it also occurs when the placeholder decides the output.

\begin{proposition}[A local zero is not the absence of flow]
\label{prop:localzero}
There are models in which the gradient and the Fisher influence of a placeholder are arbitrarily small while a finite change of the placeholder flips the output.
\end{proposition}

\begin{proof}[Proof sketch]
With two tokens of logits $z_y = \beta u$ and $z_c = 0$ at the observed $u = 1$, the gradient $\beta \sigma(-\beta)$ and the Fisher influence $\beta^2 \sigma(\beta)\sigma(-\beta)$ tend to $0$ as $\beta \to \infty$, while at $u = -1$ the greedy output switches to $c$.
For $\beta = 20$, they are $4.1 \times 10^{-8}$ and $8.2 \times 10^{-7}$, but the output still flips.
\end{proof}

\noindent
This is the shape of a short decision, a label that the placeholder determines and that the model emits with near certainty, like the first placeholder of \autoref{fig:example}.
Removal escapes parts (a) and (b), because removing a placeholder that decides the output changes the output.
It fails when the output stays likely without the placeholder, part (c), and in particular under duplicated content.
Path methods such as Integrated Gradients~\cite{pmlr-v70-sundararajan17a} behave like removal when their path starts from an input that differs only on the span, since the attributions of the span then sum to the effect of replacing the span by its starting value.

\begin{proposition}[Duplication defeats single-span removal]
\label{prop:redundancy}
If the content of a span occurs in $k \ge 2$ interchangeable copies, so that the output, including the probability that removal scores, depends on the content only through whether some copy remains, then the removal effect of every single copy is zero.
If the output depends on the content, a removal changes the output exactly when it removes every copy.
\end{proposition}

\begin{proof}[Proof sketch]
Removing one copy leaves another, so the output stays unchanged, and only a removal that contains every copy changes it.
\end{proof}

\noindent
This is the classic failure of but-for tests under overdetermination~\cite{halpern2005causes}.
Duplicated content arises when the same content enters a prompt more than once, through two placeholders or through the template.
\S\ref{sec:rq2} tests it with controlled copies.
Reading has no $(1-p)$ factor, and whether copies dilute it is an empirical question that \S\ref{sec:rq2} answers.

\begin{proposition}[Reading has no $(1-p)$ factor]
\label{prop:noFactor}
Under teacher forcing, the reading masses are functions of the attention weights of layers $1,\dots,L$, which depend only on the prompt and the recorded output tokens.
They do not depend on the final normalization, the unembedding, or the temperature of the logits.
Replacing $z$ by $z/T$ and letting $T \to 0$ drives the gradient and Fisher bounds of Proposition~\ref{prop:saturation} to zero whenever the recorded token has the unique highest logit, while leaving reading unchanged.
\end{proposition}

\begin{proof}[Proof sketch]
Reading is computed before the final normalization, and with a logit gap $g > 0$, $1 - p_y$ decays like $e^{-g/T}$, faster than the factors that rescaling adds to $J$.
\end{proof}

\begin{corollary}[Collapse forces the threshold down]
\label{cor:threshold}
Calibrate a score as in \autoref{thm:conformal}, with independent clusters.
If, with probability $\gamma > \alpha$, the true flow drawn from a cluster has a score of at most $\varepsilon_0$, then $\Pr[\tau_\alpha \le \varepsilon_0] \to 1$ as $n_c \to \infty$, and on this event every pruned placeholder has a score below $\varepsilon_0$.
\end{corollary}

\begin{proof}[Proof sketch]
The number of calibration scores at most $\varepsilon_0$ is binomial with parameters $n_c$ and $\gamma > \alpha$, so with probability tending to one it reaches $k = \lfloor \alpha(n_c+1) \rfloor$, which puts the $k$-th smallest score at or below $\varepsilon_0$.
\end{proof}

\noindent
For gradient and Fisher signals, $\varepsilon_0$ shrinks with $1-p$ by Proposition~\ref{prop:saturation}.
Whenever confident true flows make up more than an $\alpha$ fraction of the flows, the threshold falls below them, and the signal can prune only non-flows with even smaller scores.
Reading has no such factor, so confident true flows do not force its threshold toward zero.
These predictions select the hard cases of \S\ref{sec:rq2} and are tested in \S\ref{sec:rq3}.

\subsection{Hybrid Models}
\label{sec:hybrid}

P1 rests on attention being the only operation that mixes positions.
Hybrid models break this premise.
In a Mamba2 layer~\cite{dao2024ssd}, a head computes $o_t = \sum_{s \le t} K_{ts} x_s$, plus a skip term at position $t$ itself, with $K_{ts} = (C_t^{\top} B_s) \exp\!\big(A \sum_{r=s+1}^{t} \Delta_r\big) \Delta_s$~\cite{ali-etal-2025-hidden}, where $A < 0$ and the step sizes $\Delta_r$ depend on the input, as in the selective state spaces of Mamba~\cite{gu2024mamba}.

\begin{proposition}[Two flows through a state-space layer]
\label{prop:ssm}
Let the placeholder span be a single position $s^\ast < t$.
(1) Removing the write of $s^\ast$ changes $o_t$ by exactly $K_{ts^\ast} x_{s^\ast}$, a flow through the routing weight $K_{ts^\ast}$.
(2) Removing only the decay at $s^\ast$ changes $o_t$ by $(e^{F} - 1) \sum_{s < s^\ast} K_{ts} x_s$ with $F = |A|\, \Delta_{s^\ast}$, an implicit flow that does not depend on how much $s^\ast$ is read.
\end{proposition}

\begin{proof}[Proof sketch]
Part (1) follows from linearity in the writes, and part (2) from the factor $\exp(A \Delta_{s^\ast})$ that every earlier contribution passes through.
\end{proof}

\noindent
The second flow is the state-space analogue of an implicit flow through a control dependence~\cite{denning1976lattice}, since the placeholder can reset the state without being read.
Attention has no such channel to first order, because the implicit influence through the softmax normalization is itself bounded by $a_{ts}$ (Lemma~\ref{lem:head}).
Reading the attention layers of a hybrid model therefore bounds the flows that pass through attention, and these are the layers that \toolname reads (\S\ref{sec:impl}).

\subsection{Mechanization}
\label{sec:mechanization}

\paragraph{Coverage}
We mechanized every result of this section in Lean~4 with Mathlib, in 2,266 lines and 12 modules.
Each result has one entry point in the namespace \leandecl{Sightline.Paper}, which states all parts of the result as above.
Every entry point depends only on the three standard axioms of Lean (\texttt{propext}, \texttt{Classical.choice} and \texttt{Quot.sound}), and no proof uses \texttt{sorry}, the Lean command that admits a goal without proof.

\paragraph{Modeling}
A head is modeled by continuous linear maps and the softmax of \autoref{eq:attention}.
A causal pre-norm layer adds to the stream at each position $t$ the outputs of heads that read the normalized streams at the positions $i \le t$, and then applies the residual-plus-MLP map at each position.
For every stack of such layers whose components are differentiable along the forward pass, Lean derives the hypotheses of Theorem~\ref{thm:route}.
Assumption~\ref{asm:layers} thus reduces to this differentiability, which smooth activations and normalizations provide.
Corollary~\ref{cor:gif} is derived from the Lean theorem of GIF itself, so one mechanized proof joins the routing bound and the information measure of GIF.
Assumption~\ref{asm:clusters} becomes the invariance of the joint law of the clusters under permutations, which independent and identically distributed clusters satisfy, so the assumption is not vacuous, and a lemma derives from it the exchangeability of the scores drawn independently within the clusters.

\paragraph{Findings}
Mechanizing the statements made four of them more precise, since Theorem~\ref{thm:conformal} needs no random tie-breaking, Proposition~\ref{prop:noFactor} needs a unique top token, Corollary~\ref{cor:threshold} needs independent clusters, and Theorem~\ref{thm:route} also yields a zero Jacobian without a route.

\section{The \texorpdfstring{\toolname}{Sightline} Approach}
\label{sec:approach}

This section turns the properties of \S\ref{sec:theory} into \toolname, a summary that can be computed for every recorded call.

\subsection{Overview}
\label{sec:overview}

\begin{figure}[t]
\centering
\includegraphics[width=\linewidth]{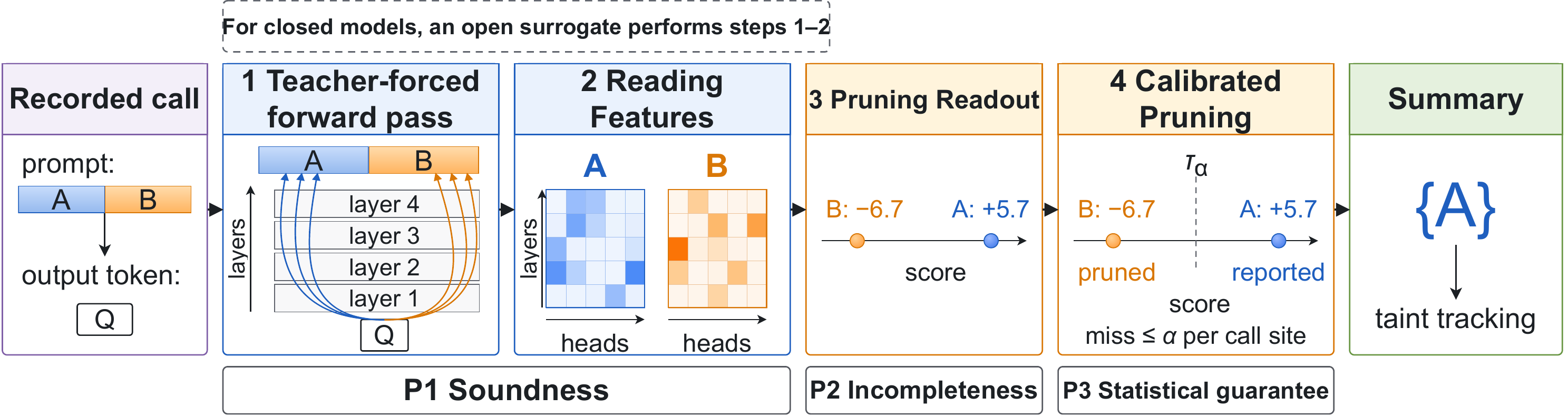}
\Description{A four-step pipeline from one teacher-forced forward pass to per-head reading masses, a readout that scores each placeholder, and a calibrated threshold that reports A and prunes B.}
\caption{Overview of \toolname on the recorded call of \autoref{fig:example}. One forward pass records the per-head reading masses, the readout scores each placeholder, and a threshold calibrated per call site reports A and prunes B.}
\label{fig:overview}
\end{figure}

\toolname computes the summary of a recorded call from a single forward pass of the model.
\autoref{fig:overview} shows its four steps.
First, it runs the model once on the recorded prompt and output with teacher forcing.
Second, it records, for every head and every placeholder, how much the output-predicting positions read the placeholder (\S\ref{sec:features}).
Third, a readout learned on these features scores each placeholder (\S\ref{sec:readout}).
Fourth, a threshold calibrated per call site reports the placeholders whose scores clear it, so that on average at most an $\alpha$ fraction of true flows is missed (\S\ref{sec:release}).
P1 motivates the evidence that the first two steps record, P2 calls for the readout of the third, and P3 supplies the guarantee of the fourth.
For models whose internals are unavailable, an open surrogate model performs the first two steps (\S\ref{sec:surrogate}).

\subsection{Reading Features}
\label{sec:features}

\begin{figure}[t]
\centering
\includegraphics[width=\linewidth]{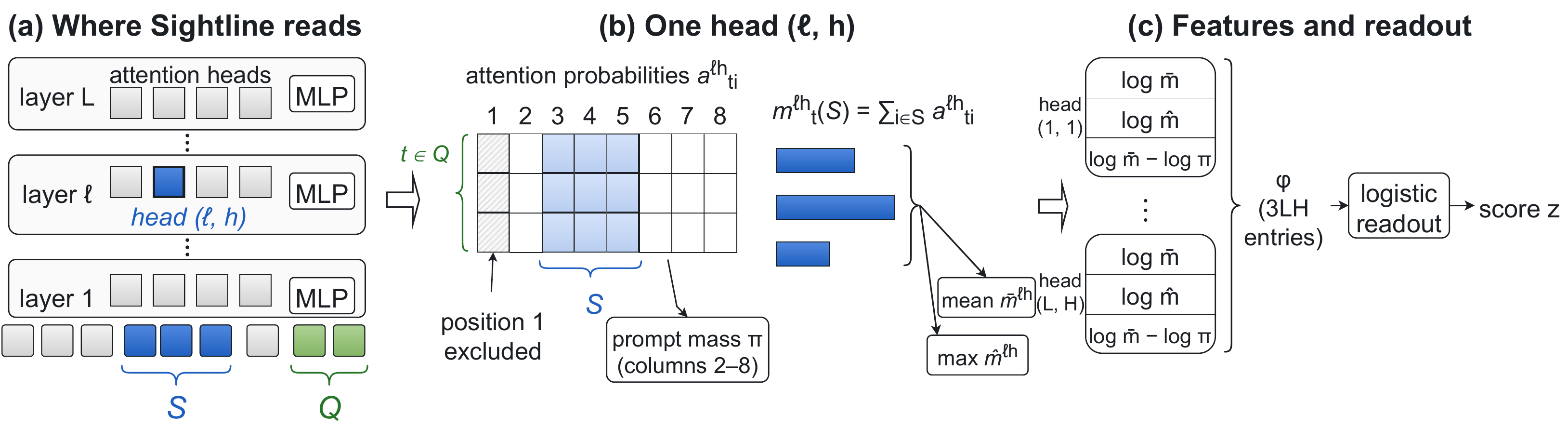}
\Description{Three panels showing which positions the output reads in a forward pass, the three statistics of one head, and the feature vector that the readout takes.}
\caption{Where \toolname reads in one teacher-forced forward pass (a), the three statistics of one head (b), and the feature vector with the readout (c), as defined in \autoref{eq:features} and \autoref{eq:vector}.}
\label{fig:features}
\end{figure}

The features measure the direct route of \autoref{thm:route} in every head (challenge \cone), since on Qwen3-8B this route carries most of the flow (\S\ref{sec:p1}).
For head $(\ell,h)$, placeholder span $\Src$ and the $n$ output-predicting positions $Q$, we compute three statistics of the reading mass $m^{\ell h}_t(\Src)$ of \autoref{eq:mass},
\begin{equation}
\label{eq:features}
\bar{m}^{\ell h} = \frac{1}{n} \sum_{t \in Q} m^{\ell h}_t(\Src),
\qquad
\hat{m}^{\ell h} = \max_{t \in Q} m^{\ell h}_t(\Src),
\qquad
\pi^{\ell h} = \frac{1}{n} \sum_{t \in Q} \sum_{j=2}^{P} a^{\ell h}_{tj} .
\end{equation}
The mean $\bar{m}$ measures how much the output reads the placeholder overall, and the peak $\hat{m}$ captures a single strong read, which is how a short decision can read its placeholder.
The prompt mass $\pi$ excludes the first position, which in many heads absorbs attention regardless of content~\cite{ICLR2024_5e5fd18f}, so that $\bar{m}/\pi$ is the share of the placeholder in what the head reads from the prompt.
The feature vector concatenates, over all layers and heads,
\begin{equation}
\label{eq:vector}
\phi = \big[\, \log(\bar{m}^{\ell h}+\epsilon),\; \log(\hat{m}^{\ell h}+\epsilon),\; \log(\bar{m}^{\ell h}+\epsilon) - \log(\pi^{\ell h}+\epsilon) \,\big]_{\ell,h},
\qquad \epsilon = 10^{-7},
\end{equation}
which has $3LH$ entries, for example $3 \times 36 \times 32 = 3{,}456$ for Qwen3-8B (\autoref{fig:features}).

We keep heads separate rather than averaging them, because P2 says that whether a read is a use depends on what the head does with it.
Averaging over heads discards exactly the information that the readout needs, and \S\ref{sec:rq1} measures the difference.
The features keep the attention weights of the direct route, since weighting the reading by the norm of the transformed values, as norm-based analyses do~\cite{kobayashi-etal-2020-attention}, does not improve Macro-F1 on Qwen3-8B.
On the call of \autoref{fig:example}, the output-predicting positions read both the query A and the delimiter B (\autoref{fig:overview}).

\subsection{Pruning Readout}
\label{sec:readout}

The readout should prune the unused reads that P2 predicts (challenge \ctwo).
We use an $\ell_2$-regularized logistic regression on standardized features, $z = w^{\top} \tilde{\phi} + b$, where $\tilde{\phi}$ subtracts the training mean and divides by the training standard deviation.
We fit it once on the reference labels of the training call sites, choose the regularization strength by five-fold cross-validation grouped by project, and then freeze it.
The readout decides from the same forward pass that produced the reading, so it adds no LLM call.
Its weights can be negative, so it is not monotone in the reading masses, which Proposition~\ref{prop:monotone} shows is needed to prune unused reads.
For a point decision, \toolname reports a placeholder as flowing when $z \ge \tau$, with $\tau$ chosen to maximize Macro-F1 on the development split.
On the call of \autoref{fig:example}, the readout scores the query A at $+5.7$ and the delimiter B at $-6.7$.

\subsection{Calibrated Pruning}
\label{sec:release}

A faint signal can hide a true flow (challenge \cthree), so when an analysis needs a bound on the flows it misses, \toolname replaces the point threshold with the threshold of \autoref{thm:conformal}.
The calibration clusters are repository families with reference labels.
From each family we draw, at random, one of its call sites with a true flow and then one of the true flows of that call site.
We take the $k$-th smallest readout score of these flows, with $k = \lfloor \alpha(n_c+1) \rfloor$, as $\tau_\alpha$, report a placeholder as flowing when $z \ge \tau_\alpha$, and prune the others.
The calibration applies to any score, so the readout affects how much the summary prunes, which \S\ref{sec:rq3} measures.
On the call of \autoref{fig:example}, the calibrated threshold lies between the two scores, so \toolname reports A and prunes B, as the reference label requires.

\subsection{Surrogate Reading for Closed Models}
\label{sec:surrogate}

Closed models expose no attention, but their calls need summaries too.
\toolname instead lets an open \emph{surrogate} model read the call of the closed model.
The surrogate teacher-forces the recorded prompt and the recorded output of the closed model, and \toolname reads the attention of the surrogate.
P1 holds for the computation of the surrogate over these tokens.
The surrogate applies its own readout and threshold, so no label of the closed model is needed, and \S\ref{sec:rq5} measures how well it predicts the flows of the closed model.

\subsection{Implementation}
\label{sec:impl}

\toolname is implemented in PyTorch~\cite{NEURIPS2019_bdbca288} on top of Hugging Face Transformers~\cite{wolf-etal-2020-transformers}.
It recomputes the attention probabilities of the output-predicting positions in strict 32-bit floating point (FP32), with the reduced-precision TF32 format disabled, so that near-zero reading masses are not rounded away.
Offline, it reads a recorded call with one forward pass.
Online, it records the same statistics while the model decodes, and in our checks the online features match the offline ones to within $3.1 \times 10^{-6}$.
For hybrid models, \toolname reads the attention layers only.
On Nemotron~3~Nano~4B~\cite{nvidia2026nemotron3nano}, these are 4 of the 42 layers, which gives 480 features.

\section{Evaluation}
\label{sec:evaluation}

Each research question (RQ) tests one property of \S\ref{sec:theory} or the practicality of the summary.
\begin{description}
\item[RQ1 (Accuracy, P1 and P2)] How accurately do reading summaries decide whether a placeholder reaches the recorded output, compared with the closest white-box methods?
\item[RQ2 (Robustness on hard cases, P4)] Do removal signals degrade on duplicated content and gradient and Fisher signals on confident outputs, as P4 predicts, while reading does not?
\item[RQ3 (Pruning under calibration, P3)] Under the same per-call-site calibration, how many spurious flows can each signal prune?
\item[RQ4 (Cost)] How much does a reading summary cost per call, offline and online?
\item[RQ5 (Transfer to closed models)] Can an open surrogate produce reading summaries for a model whose internals are unavailable?
\item[RQ6 (Use in information-flow control)] Does a reading summary help the information-flow control of an LLM agent remove the labels of inputs that do not flow while keeping those that tie executed attacks to injected content?
\end{description}
RQ1 tests how \toolname meets challenges \cone and \ctwo, and RQ3 tests challenge \cthree.
RQ2 tests the explanation of the theory for why the existing signals fail, and RQ4 to RQ6 test whether the summary is practical.

\subsection{Experimental Setup}
\label{sec:setup}

\paragraph{Data}
A summary must work on the prompts that programmers write and on programs it has never seen, so we evaluate on real call sites from open-source repositories and hold out entire repository families.
\autoref{tab:data} summarizes the two batches.
The old batch contains 8,105 placeholder--output pairs from 3,778 call sites in open-source Python projects, derived from the call sites that prior work collected from PromptSet~\cite{pister2024promptset,xu2026prism}, and we split it by project into training, development and test splits.
The new batch contains 4,918 pairs per model from 2,842 call sites in 2,103 repository families.
It comes from PromptSet repositories that the old batch does not use and from current public GitHub applications, and it shares no repository or source file with the old batch.
It keeps at most 12 call sites per family and 3 per normalized template, and its splits separate families, files and normalized templates.
Only 1.7\% of the 1,864 new calls long enough to compare have a near-duplicate template in the old batch, with a Jaccard similarity of word 5-grams of 0.5 or more after the values are removed.
Real run-time values cannot be recorded at scale, because each application needs its own keys, services or users, and many placeholders, such as user queries, retrieved documents and earlier LLM outputs, have no single true value.
We therefore reconstruct the values from code without executing it.
An LLM proposes one plausible value per placeholder from the surrounding code, prefers examples and constants that the code provides, and is instructed not to steer a value toward either label.
Three authors independently judged 95 of 100 reconstructed values plausible and disagreed on the other 5 (raw agreement 96.7\%, Gwet's AC1~\cite{Gwet_2008} of 0.97), and re-inferring 100 values with other models and relabeling the regenerated outputs kept reach for 96--98\% of the pairs.
Every readout, point threshold and trained baseline is fitted on the old batch only.

\begin{table}[t]
\caption{The two data batches. Counts of true flows and non-flows are for Qwen3-8B.}
\label{tab:data}
\small
\begin{tabular}{@{}lrrrl@{}}
\toprule
Batch & Call sites & Pairs & True flows / non-flows & Splits (pairs) \\
\midrule
Old & 3,778 & 8,105 & 7,410 / 695 & train 4,781, dev 1,676, test 1,648 \\
New & 2,842 & 4,918 & 4,336 / 582 & train 3,426, dev 659, test 833 \\
\bottomrule
\end{tabular}
\end{table}

\paragraph{Models}
White-box reading needs open weights, and P1 makes a prediction that differs by architecture, so we choose models that vary along both lines.
Qwen3-8B~\cite{yang2025qwen3}, Granite~4.2~3B~\cite{ibmgranite2026granite} and Gemma~4~E2B~\cite{google2026gemma4} are decoder-only transformers of three families and sizes, and Gemma also uses sliding-window attention in most layers.
Nemotron~3~Nano~4B~\cite{nvidia2026nemotron3nano} is a hybrid in which 21 of the 42 layers are Mamba2 layers and only 4 are attention layers.
Each model generates its own outputs for all prompts with greedy decoding and thinking disabled, from pinned checkpoint revisions.
For Qwen3-8B, generation and reading use the same 8-bit weight conversion with FP32 activations.

\paragraph{Reference labels}
Black-box per-call labeling, which decides reach as Definition~\ref{def:reach} states it and whose cost motivates this work, provides the reference.
We follow the labeling procedure of prior work~\cite{xu2026prism} with three LLM labelers from independent model families (GPT-6 Astra, Claude Opus 5.5 and DeepSeek-V4.1), which read the complete prompt, the designated placeholder and the recorded output and decide with fixed rules whether the output reflects the placeholder.
Each labels every pair independently and then, for up to three rounds, sees the labels and justifications of the other two, and the label that at least two of them endorse is the reference.
On their independent labels before discussion, the three labelers reach a Fleiss' $\kappa$~\cite{Fleiss_1971} of 0.76, a substantial agreement~\cite{Landis_1977}, and on 200 pairs sampled at random from the new batch that two authors also label independently, all five reach 0.74.
The planted values of RQ1 also prove flows without labels.

\paragraph{Baselines}
We compare against the seven methods closest to our task, namely methods proposed to decide which part of the input an LLM output used and methods proposed to control information flow through LLMs (\autoref{tab:baselines}).
Together they cover the three kinds of evidence that a white-box method can use, namely removal or replacement of the placeholder, gradient and Fisher sensitivity, and attention.
Replacement applies the mutation-based taint inference of GREYONE~\cite{gan2020greyone} to LLM calls.
We do not run TracLLM~\cite{wang2025tracllm} separately, because it targets long contexts and AttnTrace already compares against it.
ContextCite uses its original logit-scaled probability target, and we report GIF both in its original next-token form and summed over the first 32 output positions, our adaptation, which is stronger on every model.
All methods run with the same code, hyperparameters and seeds on every model, in strict FP32.
Every point threshold maximizes Macro-F1 on the old development split, and RTBAS and AT2 train their learned parts on the old training split.
Architecture forces two adjustments.
On Nemotron, attention-based methods use the 4 attention layers, and GIF on Gemma differentiates through its per-layer input embeddings as well.

\begin{table}[t]
\caption{The seven baselines closest to our task.}
\label{tab:baselines}
\small
\begin{tabular}{@{}llp{8.2cm}@{}}
\toprule
Method & Evidence & Score of a placeholder \\
\midrule
LOO & removal & drop in the log-probability of the recorded output when the placeholder is removed (on Nemotron, also from the state-space path), following the leave-one-out baselines of~\cite{cohenwang2024contextcite,wang2026attntrace} \\
ContextCite~\cite{cohenwang2024contextcite} & removal & Lasso weight of a linear model fitted to the logit-scaled output probability under 32 random removals \\
Replacement~\cite{gan2020greyone} & replacement & change in the log-probability of the recorded output when the value is replaced by an LLM-written alternative \\
GIF~\cite{storek2026gif} & Fisher & trace of the Fisher-weighted Jacobian of the output with respect to the placeholder (next token, or sum over 32 output positions) \\
RTBAS~\cite{zhong2025rtbas} & attention $\times$ gradient & long short-term memory (LSTM) network over region features of gradient-weighted attention, trained on our training labels \\
AttnTrace~\cite{wang2026attntrace} & attention & mean of the top-5 attention scores of the placeholder, averaged over 30 context subsamples ($\rho = 0.4$) for calls with at least three placeholders (our choice) \\
AT2~\cite{cohenwang2025at2} & attention & per-head weights on the reading mass, trained to predict removal effects \\
\bottomrule
\end{tabular}
\end{table}

\paragraph{Metrics}
Each RQ uses the metric that its property constrains.
For accuracy (RQ1, RQ5) we use Macro-F1, the mean of the F1 scores of the two classes, because non-flows are the minority (12--14\% of the new-batch pairs) and plain accuracy would reward reporting everything.
On our data, conservative merge, which reports everything, scores about 47.
For RQ2 we use AUROC on the cases that P4 marks as hard, because a single threshold cannot be fair to every stratum.
AUROC is the probability that a randomly chosen true flow scores above a randomly chosen non-flow, so 50 is chance.
For RQ3 we report the share of spurious flows pruned at a calibrated miss level of $\alpha = 1\%$ per call site, so that every signal is compared under the same calibration.
For RQ4 we report seconds per call.
Confidence intervals are 95\% intervals from a paired bootstrap over repository families, and we call a difference significant when its interval excludes zero.

\subsection{RQ1: Accuracy}
\label{sec:rq1}

\paragraph{Setup}
For each model we train the readout of \toolname on the old training split, select the threshold on the old development split, and score the new batch.
To check P1 on flows that do not depend on reference labels, we built 150 prompts per model that contain a unique value (a name, word, number or URL) in a placeholder.
An output that contains the value verbatim proves a flow, which holds for 113, 137, 100 and 126 outputs of Qwen3-8B, Granite, Gemma and Nemotron.

\begin{table}[t]
\caption{Macro-F1 (\%) of RQ1 on the new batch, with two controls in the last rows. $^{\circ}$ marks a difference to \toolname that is not significant.}
\label{tab:rq1}
\small
\begin{tabular}{@{}lrrrr@{}}
\toprule
Method & Qwen & Granite & Gemma & Nemotron \\
\midrule
\toolname & \textbf{86.0} & \textbf{84.1} & \textbf{86.7} & \textbf{81.1} \\
\midrule
Replacement & 80.2 & 79.6 & 80.5 & 78.7$^{\circ}$ \\
LOO & 77.5 & 77.5 & 78.8 & 75.2 \\
ContextCite & 76.6 & 77.2 & 78.1 & 79.1$^{\circ}$ \\
RTBAS & 72.9 & 73.3 & 72.9 & 74.1 \\
AT2 & 71.6 & 71.3 & 75.9 & 75.3 \\
AttnTrace & 65.5 & 65.2 & 66.2 & 70.3 \\
GIF (32 tokens) & 59.6 & 65.6 & 62.6 & 67.8 \\
GIF (next token) & 50.5 & 50.2 & 47.0 & 61.3 \\
\midrule
Hidden-state probe & 70.0 & 72.1 & 72.0 & 71.5 \\
Zero-shot labeler & 49.8 & 49.5 & 51.3 & 49.7 \\
\bottomrule
\end{tabular}
\end{table}

\paragraph{Results}
\toolname is the most accurate summary on all four models (\autoref{tab:rq1}).
On the three transformers, it leads the best baseline, Replacement, by 5.8 [3.3, 8.6], 4.5 [1.7, 7.8] and 6.2 [3.6, 9.1] points on Qwen3-8B, Granite and Gemma, and every other difference is significant.
On the hybrid Nemotron, it also has the highest score.
The threshold-free ranking agrees, with the highest AUROC on every model (93.4--96.8).

Consistent with P1, the planted values show that the whole summary, readout included, reports flows that are objectively present (challenge \cone).
\toolname reports all 476 proven flows on the four models, both at its point threshold and at the $\alpha = 1\%$ threshold of \S\ref{sec:release}, so on each model its miss rate on such flows is below 3\% at 95\% confidence.
AttnTrace and RTBAS find only 34--76\% and 42--77\% of them at their point thresholds.

The readout supplies the precision that reading alone lacks (P2, challenge \ctwo), and it draws this precision from the per-head reading.
We give every baseline the same labels and four call features (the lengths of the output, the placeholder and the prompt, and the number of placeholders) in a logistic readout trained and thresholded like ours.
All seven stay significantly below \toolname on all four models, by 3.2 to 23.8 points, and the call features alone reach 64.0--65.6.
A hidden-state probe pools the hidden states of the same forward pass at the middle and last layers over the placeholder and over the output-predicting positions, adds their elementwise product, and is trained like our readout.
It reaches 70.0--72.1, significantly below \toolname by 9.7 to 16.0 points.
Keeping any single one of the three statistics, or dropping any one of them, changes Macro-F1 by at most 1.1 points on Qwen3-8B, Granite and Nemotron, with intervals that include zero, and by at most 2.3 points on Gemma, whose layers have only 8 heads.
Averaging the heads of each layer instead costs 6.6 to 13.7 points on every model.
AT2 weighs the mean $\bar{m}$ of every head on a linear scale, and fitted to our labels it reaches only 75.8--81.0, so the signal lies in per-head reading on a log scale.
Training on shuffled labels gives a chance-level AUROC of 46--48.
Constraining the readout to be monotone in the reading masses lowers its development AUROC from 96.5 to 91.4--92.2, which is consistent with Proposition~\ref{prop:monotone}.

A local LLM does not substitute for reading either.
Qwen3-8B, prompted zero-shot with the request that the reference labelers see, answers yes for 98.5--98.9\% of the pairs, which gives a Macro-F1 of 49.5--51.3 and an AUROC of only 59.5--66.7, against 93.4--96.8 for \toolname.

\begin{answerbox}
\textbf{Answer to RQ1.} Reading summaries are the most accurate white-box summaries on all four models, 4.5 to 6.2 Macro-F1 points above the best baseline on the three transformers, and they report all 476 objectively proven flows. Given the same labels and call features, all seven baselines and the hidden-state probe stay significantly behind, so the precision comes from the readout on reading.
\end{answerbox}

\subsection{RQ2: Robustness on Hard Cases}
\label{sec:rq2}

\paragraph{Setup}
P4 names two kinds of hard cases in advance, and we test both.
For duplicated content, we built 400 prompt groups in which zero, one or two extra copies of the original value are appended to the prompt together with one unrelated distractor span.
We kept the 1,115 of the 1,200 prompts that passed a span check, the same for all four models.
We measure the AUROC of the original span against the distractor on the groups whose outputs keep the value in all three conditions (216 to 289 groups per model), so the flow is present by Definition~\ref{def:reach}.
For confident outputs, we split pairs into fifths by the confidence of the model in its own output, measured as the mean of $1 - \pmax$ over the first 32 output positions (all positions for shorter outputs).
The cut points are fixed on the development splits of both batches.
For Qwen3-8B, we evaluate the 5,907 pairs on which no readout, threshold or cut point was fitted, of which 906 fall into the most confident fifth.
For these 906 pairs, we also regenerate the outputs under five interventions on the placeholder, namely the replacement value, the values of three other pairs, and removal.

\begin{table}[t]
\caption{AUROC (\%) of RQ2 on the hard cases of P4. $^{\circ}$ marks a difference to \toolname that is not significant.}
\label{tab:rq2}
\small
\begin{tabular}{@{}lrrrr@{\hspace{1.4em}}rrrr@{}}
\toprule
& \multicolumn{4}{c}{Duplicated content ($+2$ copies)} & \multicolumn{4}{c}{Most confident fifth} \\
\cmidrule(r){2-5}\cmidrule(l){6-9}
Method & Qwen & Granite & Gemma & Nemotron & Qwen & Granite & Gemma & Nemotron \\
\midrule
\toolname & \textbf{99.0} & \textbf{99.5} & \textbf{99.8} & \textbf{100.0} & \textbf{96.0} & \textbf{94.0} & \textbf{92.7} & \textbf{90.9} \\
\midrule
Replacement & 91.8 & 88.3 & 90.7 & 90.4 & 87.3 & 90.1 & 89.0 & 88.3$^{\circ}$ \\
LOO & 67.1 & 54.0 & 68.8 & 62.0 & 84.2 & 88.9 & 88.2 & 87.0$^{\circ}$ \\
ContextCite & 90.4 & 64.9 & 88.1 & 81.5 & 89.7 & 89.4 & 89.7$^{\circ}$ & 90.2$^{\circ}$ \\
RTBAS & 93.0 & 84.2 & 95.7 & 98.7 & 83.2 & 88.5 & 82.8 & 83.4 \\
AT2 & 95.5 & 88.2 & 96.7 & 99.6 & 88.7 & 87.9 & 89.4$^{\circ}$ & 87.9$^{\circ}$ \\
AttnTrace & 70.6 & 79.3 & 73.5 & 95.9 & 86.3 & 78.6 & 79.9 & 78.6 \\
GIF (32 tokens) & 73.3 & 64.8 & 73.7 & 62.6 & 62.5 & 75.3 & 69.2 & 66.7 \\
GIF (next token) & 56.2 & 41.7 & 56.7 & 43.7 & 49.2 & 65.0 & 56.8 & 59.4 \\
\bottomrule
\end{tabular}
\end{table}

\begin{figure}[t]
\centering
\includegraphics[width=\linewidth]{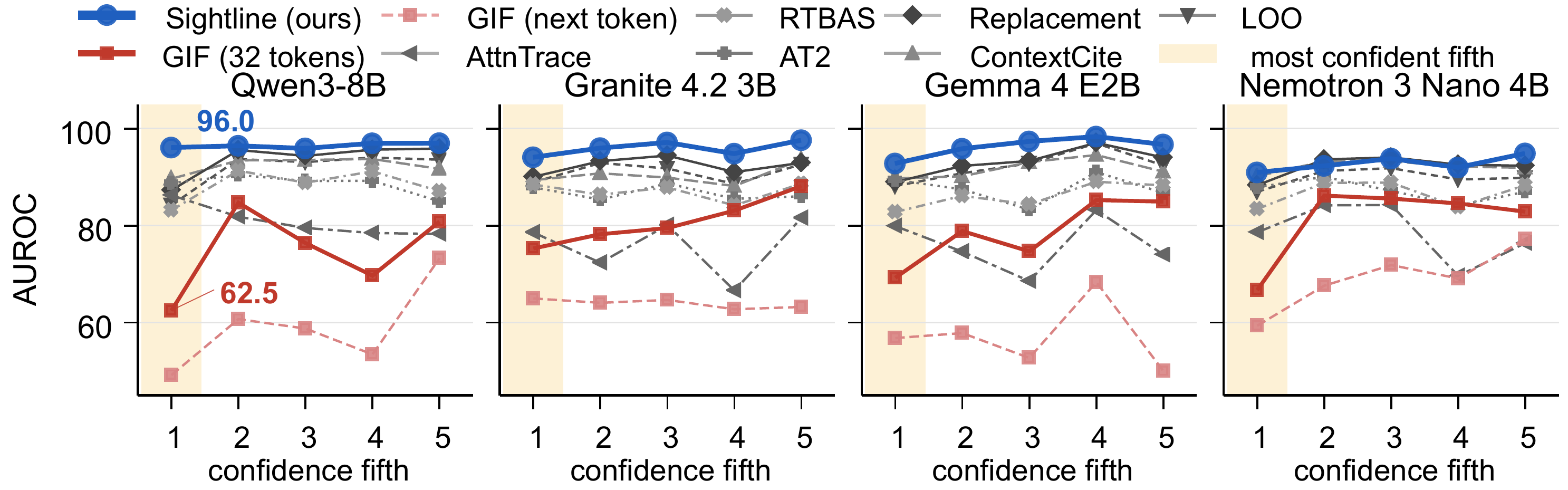}
\Description{Four line charts of AUROC by confidence fifth, one per model, in which Sightline stays flat while GIF drops in the most confident fifth.}
\caption{\toolname stays discriminative on confident outputs, while GIF drops. AUROC by confidence fifth in RQ2, one panel per model. The shaded first fifth is the most confident ($1-\pmax$ below 1\% on Qwen3-8B). Line patterns and markers distinguish the baselines.}
\label{fig:rq2}
\end{figure}

\paragraph{Results}
With duplicated content, single-span removal collapses as Proposition~\ref{prop:redundancy} predicts, and reading does not (\autoref{tab:rq2}, left).
When the value is inserted twice more, the AUROC of LOO drops from 92.9--96.6 to 54.0--68.8, and ContextCite and Replacement fall to 64.9--90.4 and 88.3--91.8.
\toolname stays at 99.0--100.0 and is significantly above every baseline on every model.

On the most confident outputs, where Proposition~\ref{prop:saturation} predicts the gradient and Fisher signals to vanish, they fall far behind (\autoref{tab:rq2}, right, and \autoref{fig:rq2}).
GIF reaches 62.5--75.3 summed over 32 tokens and 49.2--65.0 for the next token, while \toolname keeps 90.9--96.0, the highest AUROC on every model.

Reading also finds flows that a single intervention misses (challenge \cone).
Of the 84 true flows whose output the replacement value leaves unchanged but another value changes, \toolname reports 79.8\%, ContextCite 61.9\% and LOO 46.4\%.

\begin{answerbox}
\textbf{Answer to RQ2.} Where P4 predicts collapse, the predicted signals fall behind and reading does not. With the content duplicated, \toolname stays at 99.0--100.0 while LOO falls to 54.0--68.8, and on the most confident outputs, \toolname keeps the highest AUROC on every model while the Fisher signal of GIF trails it by 19 to 34 points.
\end{answerbox}

\subsection{RQ3: Pruning under Calibration}
\label{sec:rq3}

\paragraph{Setup}
An analysis that bounds the flows it misses relies on the calibration of \autoref{thm:conformal} (challenge \cthree), and the calibration applies to any score, so we give every signal the same calibration and compare how much each one prunes.
Since no method was fitted on the new batch, we calibrate and evaluate on it directly, splitting its repository families at random into a calibration half of about 1,000 families and a test half.
Each split calibrates once at $\alpha = 1\%$, as a deployment would, with the same drawn flows for every signal.
We report the means over 1,000 splits of the share of spurious flows pruned and of the miss rate per call site, the quantity of Definition~\ref{def:alpha}.
We also report the threshold that holds for the calibration set at hand, the largest $k$ for which, by the Beta law of the conformal miss rate~\cite{vovk2022alrw,angelopoulos2023gentle}, the miss rate exceeds $\alpha$ with probability at most 10\%.
To compare the rankings themselves, we further measure the pruning of every signal at the same realized miss rate of 1\%.
Across collections, we calibrate on the 152 to 155 repository families of the old development split, average over 500 draws of the calibration flows, and evaluate on the whole new batch.

\begin{table}[t]
\caption{Spurious flows pruned (\%) in RQ3 at $\alpha = 1\%$ on the new batch, calibrated on the same collection (mean of 1,000 splits) or on another collection.}
\label{tab:rq3}
\small
\begin{tabular}{@{}lrrrr@{\hspace{1.4em}}rrrr@{}}
\toprule
& \multicolumn{4}{c}{Same collection} & \multicolumn{4}{c}{Another collection} \\
\cmidrule(r){2-5}\cmidrule(l){6-9}
Method & Qwen & Granite & Gemma & Nemotron & Qwen & Granite & Gemma & Nemotron \\
\midrule
\toolname & \textbf{65.6} & \textbf{58.5} & \textbf{61.6} & \textbf{43.9} & \textbf{56.2} & \textbf{54.4} & \textbf{60.1} & \textbf{43.8} \\
\midrule
Replacement & 22.2 & 17.7 & 21.7 & 15.0 & 12.9 & 0.3 & 15.7 & 0.4 \\
LOO & 14.1 & 18.9 & 18.3 & 13.2 & 13.0 & 0.3 & 16.8 & 1.6 \\
ContextCite & 15.8 & 18.8 & 16.3 & 14.6 & 6.9 & 0.5 & 1.3 & 4.2 \\
RTBAS & 29.4 & 19.3 & 14.7 & 26.8 & 28.6 & 29.9 & 12.2 & 36.1 \\
AT2 & 15.2 & 18.1 & 37.7 & 1.1 & 7.3 & 7.2 & 38.4 & 8.1 \\
AttnTrace & 26.8 & 22.4 & 21.8 & 25.4 & 22.1 & 18.4 & 15.6 & 18.9 \\
GIF (32 tokens) & 0.6 & 0.9 & 1.3 & 12.1 & 5.5 & 22.1 & 3.1 & 21.9 \\
GIF (next token) & 3.4 & 1.6 & 0.6 & 3.6 & 0.1 & 3.2 & 0.5 & 4.2 \\
\bottomrule
\end{tabular}
\end{table}

\paragraph{Results}
Calibrated on the same collection, \toolname removes more than half of the spurious flows of conservative merge on the three transformers, pruning 65.6, 58.5, 61.6 and 43.9\% of them (\autoref{tab:rq3}, left).
This is 1.6 to 2.6 times as much as the best baseline on each model (22.4--37.7\%), while every signal misses about 1\% of true flows per call site, as \autoref{thm:conformal} guarantees for any score.
The advantage holds under the stronger guarantee and at equal misses.
With the threshold that holds for the calibration set at hand, \toolname still prunes 33.7--58.0\%, against 18.2--24.4\% for the best baseline, and at the same realized miss rate of 1\%, it prunes 45.5--65.9\%, against 22.7--39.8\%.
The controls of RQ1 prune even less under the same calibration, 18.8--24.2\% for the hidden-state probe and 3.7--6.4\% for the zero-shot labeler.
The pruning holds across kinds of placeholders.
Without the placeholders that hold code constants and those of refused calls, \toolname prunes 46.2--67.3\%, against 23.4--40.8\% for the best baseline, and on every model it prunes reconstructed values more often than code constants (48.2--68.9\% against 34.9--59.3\%).

The pruning carries over to calibration on another collection, where \toolname still prunes 56.2, 54.4, 60.1 and 43.8\% (\autoref{tab:rq3}, right) while missing about 1\% of true flows per call site.
The best baseline on each model prunes 28.6, 29.9, 38.4 and 36.1\%, and several baselines miss more under the same shift.
Every baseline prunes significantly less than \toolname.

The data show the threshold collapse of Corollary~\ref{cor:threshold} at work in the baselines.
In a separate analysis that calibrates every signal on Qwen3-8B to miss 5\% of true flows per pair, all the misses of GIF fall into the most confident fifth, where it misses 27.4\% of true flows, and it still reports 88.9\% of the spurious flows.
LOO misses 18.5\% in that fifth and reports 70.4\% of the spurious flows, whereas \toolname misses 10.1\% and reports 23.6\%.
True flows that a signal scores near zero, the confident ones for GIF and the ones whose removal leaves the output likely for LOO, pull its threshold down until it can no longer separate spurious flows.
Since the confident flows are the hardest, a client can also calibrate each confidence fifth separately, as Mondrian conformal prediction does~\cite{vovk2022alrw}.
This keeps every fifth within $\alpha$ (0.73--0.94\% for \toolname), and \toolname still prunes 42.5--61.0\%, against 21.9--27.5\% for the best baseline.

\begin{answerbox}
\textbf{Answer to RQ3.} Calibrated to miss about 1\% of true flows per call site, reading summaries prune 43.9--65.6\% of the spurious flows of unseen repositories, 1.6 to 2.6 times as many as the best baseline. The advantage holds under a guarantee for the calibration set at hand, at equal misses, per confidence fifth, and when calibrated on another collection (43.8--60.1\%). The tested sensitivity signals prune little under the calibration, as the threshold collapse of Corollary~\ref{cor:threshold} anticipates.
\end{answerbox}

\subsection{RQ4: Cost}
\label{sec:rq4}

\paragraph{Setup}
A summary is only useful if it can be computed for every call.
We time every method on the same 50 held-out calls per model, one model at a time on a single otherwise idle GPU with 48\,GB of memory, in FP32 with batch size 1, and report medians.
We also measure, for Qwen3-8B, the overhead of recording the reading statistics during generation.

\paragraph{Results}
\toolname needs one forward pass per call and takes a median of 0.12--0.20\,s per call on the four models.
The attention baselines AT2 and AttnTrace take 0.06--0.16\,s and 0.11--0.21\,s but score 5.8 to 20.5 Macro-F1 points lower (RQ1).
The removal and replacement baselines and RTBAS take 1.5 to 2.8 times as long as \toolname, because they need a forward pass per removed placeholder or random removal, or an extra backward pass.
GIF, with a forward pass plus 8 recompute and backward passes, takes 7 to 14 times as long.
Replacement also needs an LLM to write the replacement values, which took a median of 16.2\,s and about 5.4k tokens per pair with GPT-6.1 Sol, or about \$0.015 per pair.
Black-box per-call labeling, the alternative that \toolname replaces, costs about 10.4k tokens, 12.9\,s and \$0.041 per pair and labeler, and even the zero-shot labeler of RQ1 takes 0.7\,s and about 1.1k tokens per pair.
Online, recording the reading statistics while Qwen3-8B generates adds 2.1\% to the generation time at the median and 3.8\% at the 90th percentile.

\begin{answerbox}
\textbf{Answer to RQ4.} A reading summary costs one forward pass offline (0.12--0.20\,s per call), comparable to the attention baselines, a third to two thirds of the time of the removal and replacement baselines and RTBAS, and a seventh to a fourteenth of the time of GIF. Online it adds about 2\% of generation time and no API cost, whereas Replacement and black-box labeling pay for an LLM call per pair.
\end{answerbox}

\subsection{RQ5: Transfer to Closed Models}
\label{sec:rq5}

\paragraph{Setup}
Many LLM-integrated applications call closed models, which expose neither attention nor logits.
So that every model serves both as a reader and as a closed generator, we treat each of the four models in turn as closed and let each of the others read its recorded calls (\S\ref{sec:surrogate}), which gives twelve reader--generator pairs on the whole new batch.
The reader uses its own readout and threshold, fitted on its own outputs, so it needs no label of the closed model.
We compare surrogate reading with self-reading, which needs the internals, and with lexical matching (the longest common substring of placeholder and output), the natural summary when no internals are available.
To see what labels of the closed model would add, we also fit the readout of Qwen3-8B on them.

\begin{table}[t]
\caption{Macro-F1 (\%) of zero-label surrogate reading in RQ5 on the new batch. The row model reads the calls of the column model, and the diagonal is self-reading.}
\label{tab:rq5}
\small
\begin{tabular}{@{}lrrrr@{}}
\toprule
Reader $\backslash$ Generator & Qwen & Granite & Gemma & Nemotron \\
\midrule
Qwen & \emph{86.0} & 84.9 & 85.3 & 83.2 \\
Granite & 85.2 & \emph{84.1} & 84.5 & 83.4 \\
Gemma & 85.7 & 85.4 & \emph{86.7} & 84.3 \\
Nemotron & 82.4 & 82.1 & 82.5 & \emph{81.1} \\
\bottomrule
\end{tabular}
\end{table}

\paragraph{Results}
Surrogate reading needs no label of the closed model (\autoref{tab:rq5}).
With its own readout and threshold, Qwen3-8B reads the calls of Granite, Gemma and Nemotron at 84.9, 85.3 and 83.2 Macro-F1, within the interval of the self-reading of each model (+0.8 [$-1.0$, 2.9], $-1.4$ [$-3.3$, 0.2] and +2.0 [$-0.2$, 4.2]) and 14.3 to 15.5 points above lexical matching.
Across all twelve reader--generator pairs, surrogate reading differs from self-reading of the generator by $-4.2$ to $+3.2$ points, $-0.4$ on average.
Every transformer reads the calls of the hybrid Nemotron more accurately than Nemotron itself (83.2--84.3 against 81.1), whose reading covers only 4 of its 42 layers, so surrogate reading also serves hybrid models.
Labels of the closed model are not needed, since fitting the readout of Qwen3-8B on them does not improve it.

\begin{answerbox}
\textbf{Answer to RQ5.} An open surrogate produces summaries for closed models without any of their labels. Reading through Qwen3-8B matches self-reading within the interval on all three closed models and is 14.3 to 15.5 points above lexical matching, and across all twelve reader--generator pairs, surrogate reading stays within $-4.2$ to $+3.2$ points of self-reading.
\end{answerbox}

\subsection{RQ6: Use in Information-Flow Control}
\label{sec:rq6}

\paragraph{Setup}
Information-flow control for LLM agents labels each tool call with the labels of its inputs~\cite{costa2025fides}, and a summary lets it keep only the labels of inputs that flow~\cite{siddiqui2024permissive}.
We test this client on AgentDojo~\cite{debenedetti2024agentdojo}, the benchmark of injection attacks on which CaMeL, FIDES and GIF evaluate their defenses~\cite{debenedetti2025camel,costa2025fides,storek2026gif}.
We replay 291 tool-call turns of a Qwen3-8B agent under attack and compare all methods on a stratified random subset of 120 turns from 106 episodes, with 517 placeholder--call pairs.
Every method keeps the readout or parameters that it was fitted with on the old batch, and the agent labels serve only for calibration and evaluation.
We calibrate every method with episodes as the clusters of \S\ref{sec:release} and tool-call turns as their call sites, cross-fitted over five folds of episodes and repeated over 200 random splits, at $\alpha = 5\%$ (P3).
The \emph{label creep} consists of the labels that conservative merge attaches from inputs that do not flow, and we report the share of it that a method removes.
An \emph{attack pair} links a call that carries out the attack to the injected output it read, and we report the share of attack pairs that keep their label.
We also report two AUROCs, one that ranks the pairs against the reference labels and one that, as GIF does for trajectories, scores the flow from the injected block to detect calls that carry out the attack.
Granite and Gemma repeat the experiment as their own agents at half the scale.

\paragraph{Results}
On the Qwen3-8B agent, \toolname removes the most label creep, 50.0\% at an episode miss rate of 3.9\%, while almost all attack pairs keep their label.
AttnTrace comes closest (38.3\%), and the other six baselines remove at most 15.7\%, all significantly less.
Under an integrity policy in the style of FIDES, which blocks a side-effecting call whenever it carries an untrusted label, \toolname lets none of the 34 attack calls through, whether every tool output or only the injected ones count as untrusted.
In the second case, AT2, GIF and LOO let 0.3 to 1.0 through on average.
\toolname ranks pairs (AUROC 84.7) and detects executed attacks (84.6) significantly better than all seven baselines (59.8--75.8 and 44.7--73.6).
As their own agents, Granite and Gemma confirm the detection (93.2 and 83.5 AUROC, significantly above every baseline).

\begin{answerbox}
\textbf{Answer to RQ6.} As the summary of an information-flow client, reading removes the most label creep, significantly more than six of the seven baselines, while keeping the labels of almost all attack pairs and letting no attack call through an integrity policy. It ranks flows and detects executed attacks significantly better than every baseline on the Qwen3-8B agent, and it detects attacks best on the Granite and Gemma agents.
\end{answerbox}

\section{Threats to Validity}
\label{sec:discussion}
\label{sec:threats}

\paragraph{Construct validity}
Our reference labels follow the per-call labeling procedure of prior work~\cite{xu2026prism}, so the measured scores are relative to its rules, which three labelers from independent model families and two authors apply consistently (Fleiss' $\kappa$ of 0.74 on 200 sampled pairs).
Two design choices keep the comparisons fair to every method, namely the planted values of RQ1, which prove flows without labels, and the scoring of all methods against the same reference, with the threshold-free AUROC next to Macro-F1.
Reach, the notion a label-propagating analysis needs, also differs from the local flow of the theory, and \S\ref{sec:problem} separates the two, so each result is stated for the notion it concerns.

\paragraph{Internal validity}
We implemented all baselines ourselves, which could disadvantage them.
To prevent this, every method runs with the same code, hyperparameters, seeds and thresholding rule on every model, every backend reproduces the log-probabilities of the model to within $5.4 \times 10^{-5}$ nats per token, and GIF appears in both its original and a stronger adapted form.
All training and point-threshold selection used the old batch only, and the new batch was scored after every method was frozen, so no method was tuned on the data that ranks it.

\paragraph{External validity}
Our run-time values are reconstructed from code rather than observed in production.
The reconstruction grounds each value in the surrounding code and is instructed not to steer it toward a label, three authors judged 95 of 100 reconstructed values plausible (Gwet's AC1 of 0.97), and re-inferring the values with other models keeps reach for 96--98\% of the pairs.
The results also hold on 2,103 repository families that training never saw and on four models from three transformer families and one hybrid architecture.

\paragraph{Statistical validity}
The calibration bounds misses on average over call sites and over calibration sets, and it assumes that deployment clusters are exchangeable with the calibration clusters.
We therefore calibrate on families, the unit by which the data are split, average over 1,000 random splits, where every signal misses about $\alpha$, report a threshold that holds for the calibration set at hand, and also calibrate on repositories from another collection, where \toolname stays near $\alpha$.

\section{Related Work}
\label{sec:related}

\paragraph{Program analysis and information-flow control across LLM calls}
Taint analyses rely on summaries of framework code that they do not analyze~\cite{arzt2014flowdroid}.
Such summaries can be inferred by static analysis~\cite{arzt2016stubdroid}, and LLMs infer taint specifications~\cite{li2025iris}.
All of them summarize code, whereas the callee of an LLM call has none.
Information-flow analysis states flow from noninterference~\cite{goguen1982security} and lattices~\cite{denning1976lattice} to language-based enforcement~\cite{myers1999jflow,sabelfeld2003language} and dependence graphs~\cite{hammer2009flow,wasserrab2009pdg}, and taint inference by mutation~\cite{gan2020greyone} or neural approximation~\cite{she2020neutaint} and quantitative information-flow measurement~\cite{mccamant2008quantitative} measure flows on executions.
For LLM-integrated applications and agent programs, static analyses trace call chains and data flows into and out of LLM calls to find vulnerabilities~\cite{liu2024llmsmith,he2026taintp2x,wang2026agentflow}, and prior work frames the call itself as a data-flow summary problem that it solves with black-box labels~\cite{xu2026prism}.
\toolname adopts the notion of reach and the labeling procedure of that work for training and evaluation, and it replaces the labelers at run time by one forward pass of an open model, at a small fraction of their cost.
For agents, CaMeL separates control flow from untrusted data~\cite{debenedetti2025camel}, FIDES tracks confidentiality and integrity labels~\cite{costa2025fides}, and the LLMbda calculus proves probabilistic noninterference for agents whose values carry labels~\cite{garby2026llmbdacalculusaiagents}.
RTBAS and MindGuard screen tool-call dependencies with attention~\cite{zhong2025rtbas,wang2026mindguardintrinsicdecisioninspection}, NeuroTaint traces semantic taint through recorded agent traces~\cite{cai2026ghostagentredefininginformation}, and permissive information-flow analysis keeps only the labels of inputs that influenced an output~\cite{siddiqui2024permissive}.
\toolname computes this decision with one forward pass per call, and \S\ref{sec:rq6} uses it in exactly this role.
Static analyses also trade soundness for precision, guided by learning~\cite{Heo_2017} or by arguments for acceptable unsoundness~\cite{Livshits_2015}, and \toolname bounds the misses of such a trade statistically.
These analyses propagate labels through the model conservatively, label each call with another LLM, or screen dependencies without a bound on misses, and \toolname supplies the summary they need, with a miss guarantee per call site.

\paragraph{Attribution and white-box information flow in LLMs}
Context attribution asks which parts of the context an output relies on, by removal~\cite{cohenwang2024contextcite,wang2025tracllm} or attention~\cite{wang2026attntrace,cohenwang2025at2}.
We compare against both kinds, and single-span removal collapses on duplicated content, where reading does not (\S\ref{sec:rq2}).
GIF bounds the local flow to the next token through Fisher geometry, with a soundness proof mechanized in Lean~4~\cite{storek2026gif}.
Its Fisher signal vanishes on confident outputs, where \toolname keeps the highest AUROC (\S\ref{sec:rq2}).
Our mechanization also extends from the routing bound to the calibration of the pruning.
Attention and hidden states also serve other tasks, namely detecting contextual hallucinations~\cite{chuang2024lookback,sun2025redeep} and prompt injection~\cite{hung-etal-2025-attention}, compressing retrieved context~\cite{zhang2026sentineldecodingcontextutilization}, localizing instructions that guide an agent~\cite{gao2026whatguides} and decoding tool-call dependencies~\cite{sun2026toolcall}.
These methods detect, compress, localize or decode, whereas \toolname answers the question that a program analysis asks, namely which placeholders reach the output, from the direct route that P1 motivates and with a miss rate calibrated per call site.
Interpretability work aggregates attention across layers~\cite{abnar2020quantifying}, weights it by value norms~\cite{kobayashi-etal-2020-attention}, traces information-flow routes~\cite{ferrando2024routes} and approximates component effects with gradients~\cite{kramar2024atp}, and abstract interpretation certifies transformers~\cite{Bonaert_2021}.
These works estimate importance or certify robustness, whereas we prove that attention routes bound local flow and show where sensitivity provably fails.

\paragraph{Statistical guarantees}
Conformal prediction gives distribution-free, finite-sample guarantees for any score~\cite{vovk2022alrw,angelopoulos2023gentle}, with extensions to covariate shift~\cite{tibshirani2019covariate}, departures from exchangeability~\cite{barber2023beyond} and two-level hierarchical data~\cite{dunn2023hierarchical,lee2023hierarchical}.
For LLM agents, role-stratified conformal risk control calibrates detectors for tool calls per argument role~\cite{rahman2026rolecrc}.
We instead calibrate per call site, the unit at which an analysis consumes a summary, under hierarchical exchangeability and on white-box evidence, and Corollary~\ref{cor:threshold} predicts which scores can profit from the guarantee.
Because the clusters are repository families, the guarantee concerns call sites of programs that calibration never saw, and the misses in \S\ref{sec:rq3} stay near $\alpha$ on held-out families of the same collection and on another collection.

\section{Conclusion}
\label{sec:conclusion}

LLM calls leave data-flow analyses without a summary of which placeholders the output uses.
We proved in Lean~4 that, locally, a placeholder can flow to the output only along attention routes, that neither a faint reading nor faint gradient and Fisher signals of existing methods can certify that a placeholder does not flow, and that pruning therefore needs a statistical guarantee.
All thirteen results are mechanized in 2,266 lines and 12 modules of Lean, and each depends only on its three standard axioms.
\toolname turns these properties into a summary that reads a recorded call in one forward pass, prunes unused reads with a readout learned from labels, and calibrates the pruning per call site.
Each part answers one challenge, since reading bounds the local flow through a callee that has no code, the readout decides which reads are uses, and calibration over repository families bounds the misses per call site.
On 2,103 unseen repository families and four models, it is more accurate than the seven closest methods, and calibrated at $\alpha = 1\%$ it misses about 1\% of true flows and prunes 43.9--65.6\% of the spurious flows.
Calibrated on another collection, it still prunes 43.8--60.1\% of them, and it reports all 476 flows that planted values prove.
It holds up where the theory predicts that the existing signals fail, on duplicated content and on confident outputs, and on the most confident ones the Fisher signal of GIF trails it by 19 to 34 points of AUROC.
Since reading needs only the recorded tokens, an open surrogate reads models treated as closed about as accurately as they read themselves, without their labels, within $-4.2$ to $+3.2$ points of self-reading.
In the information-flow control of an LLM agent, the summary removes about half of the label creep, keeps the labels of almost all attack pairs, and lets no attack call through an integrity policy.
Reading thus carries program analyses across the NL/PL boundary with an affordable, calibrated data-flow summary of every LLM call.

\clearpage
\section*{Data-Availability Statement}

The artifact contains the implementation of \toolname (recording of reading features, readouts, calibration, and surrogate reading), our implementations of the seven baselines, the recorded calls of all four models with their reference labels and data splits, the fitted readouts, and the scripts that regenerate every table and figure.
The artifact will be archived on Zenodo with a DOI, and an anonymized copy is provided as supplementary material.
We will submit it to Artifact Evaluation.
The model checkpoints are public and pinned by revision.
The reference labels were produced with commercial models and are released for reuse.

\paragraph{Mechanized proofs}
All thirteen results of \S\ref{sec:theory} are mechanized in the artifact, a Lean~4 development with 89 theorems and 32 definitions in 2,266 lines and 12 modules.
Its trusted base is the kernel of Lean 4.28.0, Mathlib v4.28.0, the definitions of the GIF development that Corollary~\ref{cor:gif} builds on, and the axioms \texttt{propext}, \texttt{Classical.choice}, and \texttt{Quot.sound}, the only axioms on which any of the 64 declarations of its theorem inventory depends.
Corollary~\ref{cor:gif} builds on the public Lean development of GIF~\cite{storek2026gif}, which the artifact references rather than redistributes.

\bibliographystyle{ACM-Reference-Format}
\bibliography{references}

@inproceedings{kobayashi-etal-2020-attention,
    title = "Attention is Not Only a Weight: Analyzing Transformers with Vector Norms",
    author = "Kobayashi, Goro  and
      Kuribayashi, Tatsuki  and
      Yokoi, Sho  and
      Inui, Kentaro",
    editor = "Webber, Bonnie  and
      Cohn, Trevor  and
      He, Yulan  and
      Liu, Yang",
    booktitle = "Proceedings of the 2020 Conference on Empirical Methods in Natural Language Processing ({EMNLP})",
    month = nov,
    year = "2020",
    address = "Online",
    publisher = "Association for Computational Linguistics",
    url = "https://aclanthology.org/2020.emnlp-main.574/",
    doi = "10.18653/v1/2020.emnlp-main.574",
    pages = "7057--7075"
}

@inproceedings{Bonaert_2021, series={PLDI ’21}, title={Fast and precise certification of transformers}, url={http://dx.doi.org/10.1145/3453483.3454056}, DOI={10.1145/3453483.3454056}, booktitle={Proceedings of the 42nd ACM SIGPLAN International Conference on Programming Language Design and Implementation}, publisher={ACM}, author={Bonaert, Gregory and Dimitrov, Dimitar I. and Baader, Maximilian and Vechev, Martin}, year={2021}, month=jun, pages={466–481}, collection={PLDI ’21} }

@inproceedings{hung-etal-2025-attention,
    title = "Attention Tracker: Detecting Prompt Injection Attacks in {LLM}s",
    author = "Hung, Kuo-Han  and
      Ko, Ching-Yun  and
      Rawat, Ambrish  and
      Chung, I-Hsin  and
      Hsu, Winston H.  and
      Chen, Pin-Yu",
    editor = "Chiruzzo, Luis  and
      Ritter, Alan  and
      Wang, Lu",
    booktitle = "Findings of the {A}ssociation for {C}omputational {L}inguistics: {NAACL} 2025",
    month = apr,
    year = "2025",
    address = "Albuquerque, New Mexico",
    publisher = "Association for Computational Linguistics",
    url = "https://aclanthology.org/2025.findings-naacl.123/",
    doi = "10.18653/v1/2025.findings-naacl.123",
    pages = "2309--2322",
    ISBN = "979-8-89176-195-7"
}

@inproceedings{Heo_2017,
  title={Machine-Learning-Guided Selectively Unsound Static Analysis},
  url={http://dx.doi.org/10.1109/ICSE.2017.54},
  DOI={10.1109/icse.2017.54},
  booktitle={2017 IEEE/ACM 39th International Conference on Software Engineering (ICSE)},
  publisher={IEEE},
  author={Heo, Kihong and Oh, Hakjoo and Yi, Kwangkeun},
  year={2017},
  month=May,
  pages={519–529}
}

@article{Livshits_2015, title={In defense of soundiness: a manifesto}, volume={58}, ISSN={1557-7317}, url={http://dx.doi.org/10.1145/2644805}, DOI={10.1145/2644805}, number={2}, journal={Communications of the ACM}, publisher={Association for Computing Machinery (ACM)}, author={Livshits, Benjamin and Sridharan, Manu and Smaragdakis, Yannis and Lhoták, Ondřej and Amaral, J. Nelson and Chang, Bor-Yuh Evan and Guyer, Samuel Z. and Khedker, Uday P. and Møller, Anders and Vardoulakis, Dimitrios}, year={2015}, month=Jan, pages={44–46} }

@misc{storek2026gif,
      title={GIF: Locally Sound Geometric Information Flow Control for LLMs},
      author={Adam Storek and Nikolaus Holzer and Zhuo Zhang and Suman Jana},
      year={2026},
      eprint={2606.23277},
      archivePrefix={arXiv},
      primaryClass={cs.AI},
      url={https://arxiv.org/abs/2606.23277},
}

@misc{zhong2025rtbas,
      title={RTBAS: Defending LLM Agents Against Prompt Injection and Privacy Leakage},
      author={Peter Yong Zhong and Siyuan Chen and Ruiqi Wang and McKenna McCall and Ben L. Titzer and Heather Miller and Phillip B. Gibbons},
      year={2025},
      eprint={2502.08966},
      archivePrefix={arXiv},
      primaryClass={cs.CR},
      url={https://arxiv.org/abs/2502.08966},
}

@misc{cohenwang2025at2,
      title={Learning to Attribute with Attention},
      author={Benjamin Cohen-Wang and Yung-Sung Chuang and Aleksander Madry},
      year={2025},
      eprint={2504.13752},
      archivePrefix={arXiv},
      primaryClass={cs.LG},
      url={https://arxiv.org/abs/2504.13752},
}

@misc{xu2026prism,
      title={Reachability Across the NL/PL Boundary: A Taxonomy-Driven Dataflow Model for LLM-Integrated Applications},
      author={Zihao Xu and Xiao Cheng and Ruijie Meng and Yuekang Li},
      year={2026},
      eprint={2603.28345},
      archivePrefix={arXiv},
      primaryClass={cs.SE},
      url={https://arxiv.org/abs/2603.28345},
}

@article{lee2023hierarchical, title={Distribution-free inference with hierarchical data}, ISSN={3069-3497}, url={http://dx.doi.org/10.1145/3786352}, DOI={10.1145/3786352}, journal={ACM Journal of Data Science}, publisher={Association for Computing Machinery (ACM)}, author={Lee, Yonghoon and Barber, Rina and Willett, Rebecca}, year={2026}, month=Mar }

@InProceedings{dao2024ssd, title = {Transformers are {SSM}s: Generalized Models and Efficient Algorithms Through Structured State Space Duality}, author = {Dao, Tri and Gu, Albert}, booktitle = {Proceedings of the 41st International Conference on Machine Learning}, pages = {10041--10071}, year = {2024}, editor = {Salakhutdinov, Ruslan and Kolter, Zico and Heller, Katherine and Weller, Adrian and Oliver, Nuria and Scarlett, Jonathan and Berkenkamp, Felix}, volume = {235}, series = {Proceedings of Machine Learning Research}, month = {21--27 Jul}, publisher = {PMLR}, url = {https://proceedings.mlr.press/v235/dao24a.html} }

@misc{gu2024mamba,
      title={Mamba: Linear-Time Sequence Modeling with Selective State Spaces}, 
      author={Albert Gu and Tri Dao},
      year={2024},
      eprint={2312.00752},
      archivePrefix={arXiv},
      primaryClass={cs.LG},
      url={https://arxiv.org/abs/2312.00752}, 
}

@misc{kramar2024atp,
      title={AtP*: An efficient and scalable method for localizing LLM behaviour to components},
      author={János Kramár and Tom Lieberum and Rohin Shah and Neel Nanda},
      year={2024},
      eprint={2403.00745},
      archivePrefix={arXiv},
      primaryClass={cs.LG},
      url={https://arxiv.org/abs/2403.00745},
}

@inproceedings{debenedetti2025camel, title={Defeating Prompt Injections by Design}, url={http://dx.doi.org/10.1109/SaTML68715.2026.00040}, DOI={10.1109/satml68715.2026.00040}, booktitle={2026 IEEE Conference on Secure and Trustworthy Machine Learning (SaTML)}, publisher={IEEE}, author={Debenedetti, Edoardo and Shumailov, Ilia and Fan, Tianqi and Hayes, Jamie and Carlini, Nicholas and Fabian, Daniel and Kern, Christoph and Shi, Chongyang and Terzis, Andreas and Tramèr, Florian}, year={2026}, month=Mar, pages={587–618} }

@misc{costa2025fides,
      title={Securing AI Agents with Information-Flow Control},
      author={Manuel Costa and Boris Köpf and Aashish Kolluri and Andrew Paverd and Mark Russinovich and Ahmed Salem and Shruti Tople and Lukas Wutschitz and Santiago Zanella-Béguelin},
      year={2025},
      eprint={2505.23643},
      archivePrefix={arXiv},
      primaryClass={cs.CR},
      url={https://arxiv.org/abs/2505.23643},
}

@misc{rahman2026rolecrc,
      title={Beyond Aggregate Risk: Role-Stratified Conformal Risk Control for LLM Tool Calls},
      author={Md Ashikur Rahman and Md Arifur Rahman and Niamul Hassan Samin and Khandaker Rifah Tasnia and Md Hasibul Amin and Sifat Rahman Ahona and Juena Ahmed Noshin},
      year={2026},
      eprint={2607.24343},
      archivePrefix={arXiv},
      primaryClass={cs.LG},
      url={https://arxiv.org/abs/2607.24343},
}

@misc{gao2026whatguides,
      title={What Guides the Agent? Adjudicating Unauthorized Behavior via Localizing Behavior-Guiding Instructions},
      author={Yichao Gao and Yumo Zhang and Yunhao Yao and Haohua Du and Puhan Luo and Ruiqi Li and Zhiqiang Wang},
      year={2026},
      eprint={2608.24022},
      archivePrefix={arXiv},
      primaryClass={cs.CR},
      url={https://arxiv.org/abs/2608.24022},
}

@misc{sun2026toolcall,
      title={Tool-Call Dependency Structure is Linearly Decodable in LLM Agent Residual Streams},
      author={Tianda Sun and Dimitar Kazakov},
      year={2026},
      eprint={2605.25310},
      archivePrefix={arXiv},
      primaryClass={cs.CL},
      url={https://arxiv.org/abs/2605.25310},
}

@misc{wang2026agentflow,
      title={AgentFlow: Building Agent Dependency Graphs for Static Analysis of Agent Programs},
      author={Shenao Wang and Xinyi Hou and Yanjie Zhao and Xiao Cheng and Haoyu Wang},
      year={2026},
      eprint={2607.01640},
      archivePrefix={arXiv},
      primaryClass={cs.SE},
      url={https://arxiv.org/abs/2607.01640},
}

@misc{yang2025qwen3,
      title={Qwen3 Technical Report},
      author={An Yang and Anfeng Li and Baosong Yang and Beichen Zhang and Binyuan Hui and Bo Zheng and Bowen Yu and Chang Gao and Chengen Huang and Chenxu Lv and Chujie Zheng and Dayiheng Liu and Fan Zhou and Fei Huang and Feng Hu and Hao Ge and Haoran Wei and Huan Lin and Jialong Tang and Jian Yang and Jianhong Tu and Jianwei Zhang and Jianxin Yang and Jiaxi Yang and Jing Zhou and Jingren Zhou and Junyang Lin and Kai Dang and Keqin Bao and Kexin Yang and Le Yu and Lianghao Deng and Mei Li and Mingfeng Xue and Mingze Li and Pei Zhang and Peng Wang and Qin Zhu and Rui Men and Ruize Gao and Shixuan Liu and Shuang Luo and Tianhao Li and Tianyi Tang and Wenbiao Yin and Xingzhang Ren and Xinyu Wang and Xinyu Zhang and Xuancheng Ren and Yang Fan and Yang Su and Yichang Zhang and Yinger Zhang and Yu Wan and Yuqiong Liu and Zekun Wang and Zeyu Cui and Zhenru Zhang and Zhipeng Zhou and Zihan Qiu},
      year={2025},
      eprint={2505.09388},
      archivePrefix={arXiv},
      primaryClass={cs.CL},
      url={https://arxiv.org/abs/2505.09388},
}

@inproceedings{li2025iris,
 author = {Li, Ziyang and Dutta, Saikat and Naik, Mayur},
 booktitle = {International Conference on Learning Representations},
 editor = {Y. Yue and A. Garg and N. Peng and F. Sha and R. Yu},
 pages = {35735--35758},
 title = {IRIS: LLM-Assisted Static Analysis for Detecting Security Vulnerabilities},
 url = {https://proceedings.iclr.cc/paper_files/paper/2025/file/582d4e27fa24168f3af1f4582655034b-Paper-Conference.pdf},
 volume = {2025},
 year = {2025}
}

@inproceedings{sun2025redeep,
 author = {Sun, Zhongxiang and Zang, Xiaoxue and Zheng, Kai and Xu, Jun and Zhang, Xiao and Yu, Weijie and Song, Yang and Li, Han},
 booktitle = {International Conference on Learning Representations},
 editor = {Y. Yue and A. Garg and N. Peng and F. Sha and R. Yu},
 pages = {50250--50279},
 title = {ReDeEP: Detecting Hallucination in Retrieval-Augmented Generation via Mechanistic Interpretability},
 url = {https://proceedings.iclr.cc/paper_files/paper/2025/file/7daf60e805e596c3bd1e843e72ea5560-Paper-Conference.pdf},
 volume = {2025},
 year = {2025}
}

@inproceedings{vaswani2017attention,
 author = {Vaswani, Ashish and Shazeer, Noam and Parmar, Niki and Uszkoreit, Jakob and Jones, Llion and Gomez, Aidan N and Kaiser, {\L}ukasz and Polosukhin, Illia},
 booktitle = {Advances in Neural Information Processing Systems},
 editor = {I. Guyon and U. Von Luxburg and S. Bengio and H. Wallach and R. Fergus and S. Vishwanathan and R. Garnett},
 pages = {},
 publisher = {Curran Associates, Inc.},
 title = {Attention is All you Need},
 url = {https://proceedings.neurips.cc/paper_files/paper/2017/file/3f5ee243547dee91fbd053c1c4a845aa-Paper.pdf},
 volume = {30},
 year = {2017}
}

@inproceedings{tibshirani2019covariate,
 author = {Tibshirani, Ryan J and Foygel Barber, Rina and Candes, Emmanuel and Ramdas, Aaditya},
 booktitle = {Advances in Neural Information Processing Systems},
 editor = {H. Wallach and H. Larochelle and A. Beygelzimer and F. d\textquotesingle Alch\'{e}-Buc and E. Fox and R. Garnett},
 pages = {},
 publisher = {Curran Associates, Inc.},
 title = {Conformal Prediction Under Covariate Shift},
 url = {https://proceedings.neurips.cc/paper_files/paper/2019/file/8fb21ee7a2207526da55a679f0332de2-Paper.pdf},
 volume = {32},
 year = {2019}
}

@inproceedings{arzt2014flowdroid, series={PLDI ’14}, title={FlowDroid: precise context, flow, field, object-sensitive and lifecycle-aware taint analysis for Android apps}, url={http://dx.doi.org/10.1145/2594291.2594299}, DOI={10.1145/2594291.2594299}, booktitle={Proceedings of the 35th ACM SIGPLAN Conference on Programming Language Design and Implementation}, publisher={ACM}, author={Arzt, Steven and Rasthofer, Siegfried and Fritz, Christian and Bodden, Eric and Bartel, Alexandre and Klein, Jacques and Le Traon, Yves and Octeau, Damien and McDaniel, Patrick}, year={2014}, month=jun, pages={259–269}, collection={PLDI ’14} }

@inproceedings{arzt2016stubdroid, series={ICSE ’16}, title={StubDroid: automatic inference of precise data-flow summaries for the {Android} framework}, url={http://dx.doi.org/10.1145/2884781.2884816}, DOI={10.1145/2884781.2884816}, booktitle={Proceedings of the 38th International Conference on Software Engineering}, publisher={ACM}, author={Arzt, Steven and Bodden, Eric}, year={2016}, month=May, pages={725–735}, collection={ICSE ’16} }

@article{sabelfeld2003language, title={Language-based information-flow security}, volume={21}, ISSN={0733-8716}, url={http://dx.doi.org/10.1109/JSAC.2002.806121}, DOI={10.1109/jsac.2002.806121}, number={1}, journal={IEEE Journal on Selected Areas in Communications}, publisher={Institute of Electrical and Electronics Engineers (IEEE)}, author={Sabelfeld, A. and Myers, A.C.}, year={2003}, month=Jan, pages={5–19} }

@inproceedings{goguen1982security, title={Security Policies and Security Models}, url={http://dx.doi.org/10.1109/SP.1982.10014}, DOI={10.1109/sp.1982.10014}, booktitle={1982 IEEE Symposium on Security and Privacy}, publisher={IEEE}, author={Goguen, J. A. and Meseguer, J.}, year={1982}, month=Apr, pages={11–20} }

@inproceedings{mccamant2008quantitative, series={PLDI ’08}, title={Quantitative information flow as network flow capacity}, url={http://dx.doi.org/10.1145/1375581.1375606}, DOI={10.1145/1375581.1375606}, booktitle={Proceedings of the 29th ACM SIGPLAN Conference on Programming Language Design and Implementation}, publisher={ACM}, author={McCamant, Stephen and Ernst, Michael D.}, year={2008}, month=jun, pages={193–205}, collection={PLDI ’08} }

@book{vovk2022alrw, title={Algorithmic Learning in a Random World}, ISBN={9783031066498}, url={http://dx.doi.org/10.1007/978-3-031-06649-8}, DOI={10.1007/978-3-031-06649-8}, publisher={Springer International Publishing}, author={Vovk, Vladimir and Gammerman, Alexander and Shafer, Glenn}, year={2022}, edition={second} }

@article{angelopoulos2023gentle, title={Conformal Prediction: A Gentle Introduction}, volume={16}, ISSN={1935-8245}, url={http://dx.doi.org/10.1561/2200000101}, DOI={10.1561/2200000101}, number={4}, journal={Foundations and Trends® in Machine Learning}, publisher={Emerald}, author={Angelopoulos, Anastasios N. and Bates, Stephen}, year={2023}, month=Mar, pages={494–591} }

@article{hammer2009flow, title={Flow-sensitive, context-sensitive, and object-sensitive information flow control based on program dependence graphs}, volume={8}, ISSN={1615-5270}, url={http://dx.doi.org/10.1007/s10207-009-0086-1}, DOI={10.1007/s10207-009-0086-1}, number={6}, journal={International Journal of Information Security}, publisher={Springer Science and Business Media LLC}, author={Hammer, Christian and Snelting, Gregor}, year={2009}, month=dec, pages={399–422} }

@article{halpern2005causes, title={Causes and Explanations: A Structural-Model Approach. Part I: Causes}, volume={56}, ISSN={1464-3537}, url={http://dx.doi.org/10.1093/bjps/axi147}, DOI={10.1093/bjps/axi147}, number={4}, journal={The British Journal for the Philosophy of Science}, publisher={University of Chicago Press}, author={Halpern, Joseph Y. and Pearl, Judea}, year={2005}, month=Dec, pages={843–887} }

@article{denning1976lattice, title={A lattice model of secure information flow}, volume={19}, ISSN={1557-7317}, url={http://dx.doi.org/10.1145/360051.360056}, DOI={10.1145/360051.360056}, number={5}, journal={Communications of the ACM}, publisher={Association for Computing Machinery (ACM)}, author={Denning, Dorothy E.}, year={1976}, month=May, pages={236–243} }

@inproceedings{wasserrab2009pdg, series={PLAS ’09}, title={On PDG-based noninterference and its modular proof}, url={http://dx.doi.org/10.1145/1554339.1554345}, DOI={10.1145/1554339.1554345}, booktitle={Proceedings of the ACM SIGPLAN Fourth Workshop on Programming Languages and Analysis for Security}, publisher={ACM}, author={Wasserrab, Daniel and Lohner, Denis and Snelting, Gregor}, year={2009}, month=jun, pages={31–44}, collection={PLAS ’09} }

@inproceedings{myers1999jflow, series={POPL99}, title={JFlow: practical mostly-static information flow control}, url={http://dx.doi.org/10.1145/292540.292561}, DOI={10.1145/292540.292561}, booktitle={Proceedings of the 26th ACM SIGPLAN-SIGACT symposium on Principles of programming languages}, publisher={ACM}, author={Myers, Andrew C.}, year={1999}, month=Jan, pages={228–241}, collection={POPL99} }

@article{dunn2023hierarchical, title={Distribution-Free Prediction Sets for Two-Layer Hierarchical Models}, volume={118}, ISSN={1537-274X}, url={http://dx.doi.org/10.1080/01621459.2022.2060112}, DOI={10.1080/01621459.2022.2060112}, number={544}, journal={Journal of the American Statistical Association}, publisher={Informa UK Limited}, author={Dunn, Robin and Wasserman, Larry and Ramdas, Aaditya}, year={2023}, pages={2491–2502} }

@inproceedings{he2026taintp2x, series={ICSE ’26}, title={TaintP2X: Detecting Taint-Style Prompt-to-Anything Injection Vulnerabilities in LLM-Integrated Applications}, url={http://dx.doi.org/10.1145/3744916.3773199}, DOI={10.1145/3744916.3773199}, booktitle={Proceedings of the 2026 IEEE/ACM 48th International Conference on Software Engineering}, publisher={ACM}, author={He, Junjie and Wang, Shenao and Zhao, Yanjie and Hou, Xinyi and Liu, Zhao and Zou, Quanchen and Wang, Haoyu}, year={2026}, month=Apr, pages={1968–1979}, collection={ICSE ’26} }

@inproceedings{liu2024llmsmith, series={CCS ’24}, title={Demystifying RCE Vulnerabilities in LLM-Integrated Apps}, url={http://dx.doi.org/10.1145/3658644.3690338}, DOI={10.1145/3658644.3690338}, booktitle={Proceedings of the 2024 on ACM SIGSAC Conference on Computer and Communications Security}, publisher={ACM}, author={Liu, Tong and Deng, Zizhuang and Meng, Guozhu and Li, Yuekang and Chen, Kai}, year={2024}, month=Dec, pages={1716–1730}, collection={CCS ’24} }

@inproceedings{pister2024promptset, series={LLM4Code ’24}, title={PromptSet: A Programmer’s Prompting Dataset}, url={http://dx.doi.org/10.1145/3643795.3648395}, DOI={10.1145/3643795.3648395}, booktitle={Proceedings of the 1st International Workshop on Large Language Models for Code}, publisher={ACM}, author={Pister, Kaiser and Paul, Dhruba Jyoti and Joshi, Ishan and Brophy, Patrick}, year={2024}, month=Apr, pages={62–69}, collection={LLM4Code ’24} }

@inproceedings{jain2019attention,
    title = "{A}ttention is not {E}xplanation",
    author = "Jain, Sarthak  and
      Wallace, Byron C.",
    editor = "Burstein, Jill  and
      Doran, Christy  and
      Solorio, Thamar",
    booktitle = "Proceedings of the 2019 Conference of the {N}orth {A}merican Chapter of the {A}ssociation for {C}omputational {L}inguistics: Human Language Technologies, Volume 1 (Long and Short Papers)",
    month = jun,
    year = "2019",
    address = "Minneapolis, Minnesota",
    publisher = "Association for Computational Linguistics",
    url = "https://aclanthology.org/N19-1357/",
    doi = "10.18653/v1/N19-1357",
    pages = "3543--3556"
}

@inproceedings{wiegreffe2019attention,
    title = "Attention is not not Explanation",
    author = "Wiegreffe, Sarah  and
      Pinter, Yuval",
    editor = "Inui, Kentaro  and
      Jiang, Jing  and
      Ng, Vincent  and
      Wan, Xiaojun",
    booktitle = "Proceedings of the 2019 Conference on Empirical Methods in Natural Language Processing and the 9th International Joint Conference on Natural Language Processing ({EMNLP}-{IJCNLP})",
    month = nov,
    year = "2019",
    address = "Hong Kong, China",
    publisher = "Association for Computational Linguistics",
    url = "https://aclanthology.org/D19-1002/",
    doi = "10.18653/v1/D19-1002",
    pages = "11--20"
}

@inproceedings{abnar2020quantifying,
    title = "Quantifying Attention Flow in Transformers",
    author = "Abnar, Samira  and
      Zuidema, Willem",
    editor = "Jurafsky, Dan  and
      Chai, Joyce  and
      Schluter, Natalie  and
      Tetreault, Joel",
    booktitle = "Proceedings of the 58th Annual Meeting of the {A}ssociation for {C}omputational {L}inguistics",
    month = jul,
    year = "2020",
    address = "Online",
    publisher = "Association for Computational Linguistics",
    url = "https://aclanthology.org/2020.acl-main.385/",
    doi = "10.18653/v1/2020.acl-main.385",
    pages = "4190--4197"
}

@inproceedings{chuang2024lookback,
    title = "Lookback Lens: Detecting and Mitigating Contextual Hallucinations in Large Language Models Using Only Attention Maps",
    author = "Chuang, Yung-Sung  and
      Qiu, Linlu  and
      Hsieh, Cheng-Yu  and
      Krishna, Ranjay  and
      Kim, Yoon  and
      Glass, James R.",
    editor = "Al-Onaizan, Yaser  and
      Bansal, Mohit  and
      Chen, Yun-Nung",
    booktitle = "Proceedings of the 2024 Conference on Empirical Methods in Natural Language Processing",
    month = nov,
    year = "2024",
    address = "Miami, Florida, USA",
    publisher = "Association for Computational Linguistics",
    url = "https://aclanthology.org/2024.emnlp-main.84/",
    doi = "10.18653/v1/2024.emnlp-main.84",
    pages = "1419--1436"
}

@inproceedings{wang2026attntrace, title={AttnTrace: Contextual Attribution of Prompt Injection and Knowledge Corruption}, url={http://dx.doi.org/10.1109/SP63933.2026.00198}, DOI={10.1109/sp63933.2026.00198}, booktitle={2026 IEEE Symposium on Security and Privacy (SP)}, publisher={IEEE}, author={Wang, Yanting and Geng, Runpeng and Chen, Ying and Jia, Jinyuan}, year={2026}, month=May, pages={4320–4338} }

@inproceedings{cohenwang2024contextcite,
 author = {Cohen-Wang, Benjamin and Shah, Harshay and Georgiev, Kristian and M\k{a}dry, Aleksander},
 booktitle = {Advances in Neural Information Processing Systems},
 doi = {10.52202/079017-3035},
 editor = {A. Globerson and L. Mackey and D. Belgrave and A. Fan and U. Paquet and J. Tomczak and C. Zhang},
 pages = {95764--95807},
 publisher = {Curran Associates, Inc.},
 title = {ContextCite: Attributing Model Generation to Context},
 url = {https://proceedings.neurips.cc/paper_files/paper/2024/file/adbea136219b64db96a9941e4249a857-Paper-Conference.pdf},
 volume = {37},
 year = {2024}
}

@inproceedings{debenedetti2024agentdojo,
 author = {Debenedetti, Edoardo and Zhang, Jie and Balunovic, Mislav and Beurer-Kellner, Luca and Fischer, Marc and Tram\`{e}r, Florian},
 booktitle = {Advances in Neural Information Processing Systems},
 doi = {10.52202/079017-2636},
 editor = {A. Globerson and L. Mackey and D. Belgrave and A. Fan and U. Paquet and J. Tomczak and C. Zhang},
 pages = {82895--82920},
 publisher = {Curran Associates, Inc.},
 title = {AgentDojo: A Dynamic Environment to Evaluate Prompt Injection Attacks and Defenses for LLM Agents},
 url = {https://proceedings.neurips.cc/paper_files/paper/2024/file/97091a5177d8dc64b1da8bf3e1f6fb54-Paper-Datasets_and_Benchmarks_Track.pdf},
 volume = {37},
 year = {2024}
}

@inproceedings{gan2020greyone,
author = {Shuitao Gan and Chao Zhang and Peng Chen and Bodong Zhao and Xiaojun Qin and Dong Wu and Zuoning Chen},
title = {{GREYONE}: Data Flow Sensitive Fuzzing},
booktitle = {29th USENIX Security Symposium (USENIX Security 20)},
year = {2020},
isbn = {978-1-939133-17-5},
pages = {2577--2594},
url = {https://www.usenix.org/conference/usenixsecurity20/presentation/gan},
publisher = {USENIX Association},
month = aug
}

@inproceedings{wang2025tracllm,
author = {Yanting Wang and Wei Zou and Runpeng Geng and Jinyuan Jia},
title = {{TracLLM}: A Generic Framework for Attributing Long Context {LLMs}},
booktitle = {34th USENIX Security Symposium (USENIX Security 25)},
year = {2025},
isbn = {978-1-939133-52-6},
address = {Seattle, WA},
pages = {3845--3864},
url = {https://www.usenix.org/conference/usenixsecurity25/presentation/wang-yanting},
publisher = {USENIX Association},
month = aug
}

@misc{ibmgranite2026granite,
  author       = {{IBM Granite}},
  title        = {{granite-4.2-3b}},
  year         = {2026},
  howpublished = {Hugging Face model repository},
  url          = {https://huggingface.co/ibm-granite/granite-4.2-3b},
  note         = {Revision e459acceac81}
}

@misc{google2026gemma4,
      title={Gemma 4 Technical Report},
      author={{Gemma Team}},
      year={2026},
      eprint={2607.02770},
      archivePrefix={arXiv},
      primaryClass={cs.CL},
      url={https://arxiv.org/abs/2607.02770},
}

@misc{nvidia2026nemotron3nano,
  author       = {{NVIDIA}},
  title        = {{NVIDIA-Nemotron-3-Nano-4B-BF16}},
  year         = {2026},
  howpublished = {Hugging Face model repository},
  url          = {https://huggingface.co/nvidia/NVIDIA-Nemotron-3-Nano-4B-BF16},
  note         = {Revision dfaf35de3e30}
}

@inproceedings{she2020neutaint, title={Neutaint: Efficient Dynamic Taint Analysis with Neural Networks}, url={http://dx.doi.org/10.1109/SP40000.2020.00022}, DOI={10.1109/sp40000.2020.00022}, booktitle={2020 IEEE Symposium on Security and Privacy (SP)}, publisher={IEEE}, author={She, Dongdong and Chen, Yizheng and Shah, Abhishek and Ray, Baishakhi and Jana, Suman}, year={2020}, month=May, pages={1527–1543} }

@article{barber2023beyond, title={Conformal prediction beyond exchangeability}, volume={51}, ISSN={2168-8966}, url={http://dx.doi.org/10.1214/23-AOS2276}, DOI={10.1214/23-aos2276}, number={2}, journal={The Annals of Statistics}, publisher={Institute of Mathematical Statistics}, author={Barber, Rina Foygel and Candès, Emmanuel J. and Ramdas, Aaditya and Tibshirani, Ryan J.}, year={2023}, month=Apr, pages={816–845} }

@article{siddiqui2024permissive,
title={Permissive Information-Flow Analysis for Large Language Models},
author={Shoaib Ahmed Siddiqui and Radhika Gaonkar and Boris K{\"o}pf and David Krueger and Andrew Paverd and Ahmed Salem and Shruti Tople and Lukas Wutschitz and Menglin Xia and Santiago Zanella-Beguelin},
journal={Transactions on Machine Learning Research},
issn={2835-8856},
year={2025},
url={https://openreview.net/forum?id=ufYRO8y3mr},
note={}
}

@inproceedings{ferrando2024routes,
    title = "Information Flow Routes: Automatically Interpreting Language Models at Scale",
    author = "Ferrando, Javier  and
      Voita, Elena",
    editor = "Al-Onaizan, Yaser  and
      Bansal, Mohit  and
      Chen, Yun-Nung",
    booktitle = "Proceedings of the 2024 Conference on Empirical Methods in Natural Language Processing",
    month = nov,
    year = "2024",
    address = "Miami, Florida, USA",
    publisher = "Association for Computational Linguistics",
    url = "https://aclanthology.org/2024.emnlp-main.965/",
    doi = "10.18653/v1/2024.emnlp-main.965",
    pages = "17432--17445"
}

@article{elhage2021framework,
   title={A Mathematical Framework for Transformer Circuits},
   author={Elhage, Nelson and Nanda, Neel and Olsson, Catherine and Henighan, Tom and Joseph, Nicholas and Mann, Ben and Askell, Amanda and Bai, Yuntao and Chen, Anna and Conerly, Tom and DasSarma, Nova and Drain, Dawn and Ganguli, Deep and Hatfield-Dodds, Zac and Hernandez, Danny and Jones, Andy and Kernion, Jackson and Lovitt, Liane and Ndousse, Kamal and Amodei, Dario and Brown, Tom and Clark, Jack and Kaplan, Jared and McCandlish, Sam and Olah, Chris},
   year={2021},
   journal={Transformer Circuits Thread},
   note={https://transformer-circuits.pub/2021/framework/index.html}
}

@article{Su_2024, title={RoFormer: Enhanced transformer with Rotary Position Embedding}, volume={568}, ISSN={0925-2312}, url={http://dx.doi.org/10.1016/j.neucom.2023.127063}, DOI={10.1016/j.neucom.2023.127063}, journal={Neurocomputing}, publisher={Elsevier BV}, author={Su, Jianlin and Ahmed, Murtadha and Lu, Yu and Pan, Shengfeng and Bo, Wen and Liu, Yunfeng}, year={2024}, month=Feb, pages={127063} }

@inproceedings{ainslie-etal-2023-gqa,
    title = "{GQA}: Training Generalized Multi-Query Transformer Models from Multi-Head Checkpoints",
    author = "Ainslie, Joshua  and
      Lee-Thorp, James  and
      de Jong, Michiel  and
      Zemlyanskiy, Yury  and
      Lebron, Federico  and
      Sanghai, Sumit",
    editor = "Bouamor, Houda  and
      Pino, Juan  and
      Bali, Kalika",
    booktitle = "Proceedings of the 2023 Conference on Empirical Methods in Natural Language Processing",
    month = dec,
    year = "2023",
    address = "Singapore",
    publisher = "Association for Computational Linguistics",
    url = "https://aclanthology.org/2023.emnlp-main.298/",
    doi = "10.18653/v1/2023.emnlp-main.298",
    pages = "4895--4901"
}

@inproceedings{ICLR2024_5e5fd18f,
 author = {Xiao, Guangxuan and Tian, Yuandong and Chen, Beidi and Han, Song and Lewis, Mike },
 booktitle = {International Conference on Learning Representations},
 editor = {B. Kim and Y. Yue and S. Chaudhuri and K. Fragkiadaki and M. Khan and Y. Sun},
 pages = {21875--21895},
 title = {Efficient Streaming Language Models with Attention Sinks},
 url = {https://proceedings.iclr.cc/paper_files/paper/2024/file/5e5fd18f863cbe6d8ae392a93fd271c9-Paper-Conference.pdf},
 volume = {2024},
 year = {2024}
}

@inproceedings{NEURIPS2019_bdbca288,
 author = {Paszke, Adam and Gross, Sam and Massa, Francisco and Lerer, Adam and Bradbury, James and Chanan, Gregory and Killeen, Trevor and Lin, Zeming and Gimelshein, Natalia and Antiga, Luca and Desmaison, Alban and Kopf, Andreas and Yang, Edward and DeVito, Zachary and Raison, Martin and Tejani, Alykhan and Chilamkurthy, Sasank and Steiner, Benoit and Fang, Lu and Bai, Junjie and Chintala, Soumith},
 booktitle = {Advances in Neural Information Processing Systems},
 editor = {H. Wallach and H. Larochelle and A. Beygelzimer and F. d\textquotesingle Alch\'{e}-Buc and E. Fox and R. Garnett},
 pages = {},
 publisher = {Curran Associates, Inc.},
 title = {PyTorch: An Imperative Style, High-Performance Deep Learning Library},
 url = {https://proceedings.neurips.cc/paper_files/paper/2019/file/bdbca288fee7f92f2bfa9f7012727740-Paper.pdf},
 volume = {32},
 year = {2019}
}

@inproceedings{wolf-etal-2020-transformers,
    title = "Transformers: State-of-the-Art Natural Language Processing",
    author = "Wolf, Thomas  and
      Debut, Lysandre  and
      Sanh, Victor  and
      Chaumond, Julien  and
      Delangue, Clement  and
      Moi, Anthony  and
      Cistac, Pierric  and
      Rault, Tim  and
      Louf, Remi  and
      Funtowicz, Morgan  and
      Davison, Joe  and
      Shleifer, Sam  and
      von Platen, Patrick  and
      Ma, Clara  and
      Jernite, Yacine  and
      Plu, Julien  and
      Xu, Canwen  and
      Le Scao, Teven  and
      Gugger, Sylvain  and
      Drame, Mariama  and
      Lhoest, Quentin  and
      Rush, Alexander",
    editor = "Liu, Qun  and
      Schlangen, David",
    booktitle = "Proceedings of the 2020 Conference on Empirical Methods in Natural Language Processing: System Demonstrations",
    month = oct,
    year = "2020",
    address = "Online",
    publisher = "Association for Computational Linguistics",
    url = "https://aclanthology.org/2020.emnlp-demos.6/",
    doi = "10.18653/v1/2020.emnlp-demos.6",
    pages = "38--45"
}

@InProceedings{demoura2021lean4,
author="Moura, Leonardo de
and Ullrich, Sebastian",
editor="Platzer, Andr{\'e}
and Sutcliffe, Geoff",
title="The Lean 4 Theorem Prover and Programming Language",
booktitle="Automated Deduction -- CADE 28",
year="2021",
publisher="Springer International Publishing",
address="Cham",
pages="625--635",
isbn="978-3-030-79876-5"
}

@inproceedings{mathlib2020, series={CPP ’20}, title={The {Lean} mathematical library}, url={http://dx.doi.org/10.1145/3372885.3373824}, DOI={10.1145/3372885.3373824}, booktitle={Proceedings of the 9th ACM SIGPLAN International Conference on Certified Programs and Proofs}, publisher={ACM}, author={The mathlib Community}, year={2020}, month=Jan, pages={367–381}, collection={CPP ’20} }

@article{Gwet_2008, title={Computing inter‐rater reliability and its variance in the presence of high agreement}, volume={61}, ISSN={2044-8317}, url={http://dx.doi.org/10.1348/000711006X126600}, DOI={10.1348/000711006x126600}, number={1}, journal={British Journal of Mathematical and Statistical Psychology}, publisher={Wiley}, author={Gwet, Kilem Li}, year={2008}, month=May, pages={29–48} }

@article{Landis_1977, title={The Measurement of Observer Agreement for Categorical Data}, volume={33}, ISSN={0006-341X}, url={http://dx.doi.org/10.2307/2529310}, DOI={10.2307/2529310}, number={1}, journal={Biometrics}, publisher={JSTOR}, author={Landis, J. Richard and Koch, Gary G.}, year={1977}, month=Mar, pages={159–174} }

@misc{garby2026llmbdacalculusaiagents,
      title={The LLMbda Calculus: AI Agents, Conversations, and Information Flow}, 
      author={Zac Garby and Andrew D. Gordon and David Sands},
      year={2026},
      eprint={2602.20064},
      archivePrefix={arXiv},
      primaryClass={cs.PL},
      url={https://arxiv.org/abs/2602.20064}, 
}

@misc{cai2026ghostagentredefininginformation,
      title={Ghost in the Agent: Redefining Information Flow Tracking for LLM Agents}, 
      author={Yuandao Cai and Wensheng Tang and Cheng Wen and Shengchao Qin},
      year={2026},
      eprint={2604.23374},
      archivePrefix={arXiv},
      primaryClass={cs.CR},
      url={https://arxiv.org/abs/2604.23374}, 
}

@misc{wang2026mindguardintrinsicdecisioninspection,
      title={MindGuard: Intrinsic Decision Inspection for Securing LLM Agents Against Metadata Poisoning}, 
      author={Zhiqiang Wang and Haohua Du and Guanquan Shi and Junyang Zhang and HaoRan Cheng and Yunhao Yao and Kaiwen Guo and Xiang-Yang Li},
      year={2026},
      eprint={2508.20412},
      archivePrefix={arXiv},
      primaryClass={cs.CR},
      url={https://arxiv.org/abs/2508.20412}, 
}

@misc{zhang2026sentineldecodingcontextutilization,
      title={Sentinel: Decoding Context Utilization via Attention Probing for Efficient LLM Context Compression}, 
      author={Yong Zhang and Heng Li and Yanwen Huang and Ning Cheng and Yang Guo and Yun Zhu and Yanmeng Wang and Shaojun Wang and Jing Xiao},
      year={2026},
      eprint={2505.23277},
      archivePrefix={arXiv},
      primaryClass={cs.CL},
      url={https://arxiv.org/abs/2505.23277}, 
}

@InProceedings{pmlr-v70-sundararajan17a,
  title = {Axiomatic Attribution for Deep Networks},
  author = {Mukund Sundararajan and Ankur Taly and Qiqi Yan},
  booktitle = {Proceedings of the 34th International Conference on Machine Learning},
  pages = {3319--3328},
  year = {2017},
  editor = {Precup, Doina and Teh, Yee Whye},
  volume = {70},
  series = {Proceedings of Machine Learning Research},
  month = {06--11 Aug},
  publisher = {PMLR},
  url = {https://proceedings.mlr.press/v70/sundararajan17a.html}
}

@inproceedings{ali-etal-2025-hidden,
    title = "The Hidden Attention of Mamba Models",
    author = "Ali, Ameen  and
      Zimerman, Itamar  and
      Wolf, Lior",
    editor = "Che, Wanxiang  and
      Nabende, Joyce  and
      Shutova, Ekaterina  and
      Pilehvar, Mohammad Taher",
    booktitle = "Proceedings of the 63rd Annual Meeting of the {A}ssociation for {C}omputational {L}inguistics (Volume 1: Long Papers)",
    month = jul,
    year = "2025",
    address = "Vienna, Austria",
    publisher = "Association for Computational Linguistics",
    url = "https://aclanthology.org/2025.acl-long.76/",
    doi = "10.18653/v1/2025.acl-long.76",
    pages = "1516--1534",
    ISBN = "979-8-89176-251-0"
}

@article{Fleiss_1971, title={Measuring nominal scale agreement among many raters.}, volume={76}, ISSN={0033-2909}, url={http://dx.doi.org/10.1037/h0031619}, DOI={10.1037/h0031619}, number={5}, journal={Psychological Bulletin}, publisher={American Psychological Association (APA)}, author={Fleiss, Joseph L.}, year={1971}, month=Nov, pages={378–382} }

@misc{beltagy2020longformerlongdocumenttransformer,
      title={Longformer: The Long-Document Transformer}, 
      author={Iz Beltagy and Matthew E. Peters and Arman Cohan},
      year={2020},
      eprint={2004.05150},
      archivePrefix={arXiv},
      primaryClass={cs.CL},
      url={https://arxiv.org/abs/2004.05150}, 
}

\end{document}